\documentclass[aps,pre,twocolumn,floatfix,superscriptaddress]{revtex4}

\usepackage{graphicx}  
\usepackage{dcolumn}   
\usepackage{bm}        
\usepackage{amssymb}   
\usepackage[latin1]{inputenc}
\usepackage{amsfonts}
\usepackage{amsmath}
\usepackage{amsthm}
\usepackage{xcolor}
\usepackage{fontenc}
\usepackage{textcomp}
\usepackage{epstopdf}
\usepackage{hyperref}
\begin{document}
\title{Herd Immunity with Spatial Adaptation Based on Global Prevalence Information}

\author{Akhil Panicker}
    \email[Correspondence email address:]{panickeratp@gmail.com}
    \affiliation{Department of Physics, Cochin University of Science and Technology, Cochin, 682022, Kerala, India}
\author{V. Sasidevan}
    \email[Correspondence email address:]{sasidevan@gmail.com}
    \affiliation{Department of Physics, Cochin University of Science and Technology, Cochin, 682022, Kerala, India}

\date{\today} 

\begin{abstract}
During an epidemic outbreak, individuals often modify their behavior in response to global prevalence cues, using spatially mediated adaptations such as reduced mobility or transmission range. In this work, we investigate the impact of distance-based adaptive behaviors on epidemic dynamics, where a fraction of the population adjusts its transmission range and susceptibility to infection based on global prevalence. We consider three adaptation scenarios: a constant adaptive fraction, a power-law dependence and a sigmoidal dependence of adaptive fraction on global prevalence. In the spatially well-mixed regime, we analytically obtain critical adaptation thresholds necessary for epidemic mitigation and in the spatially static regime, we establish bounds for the thresholds using continuum percolation results. Our results indicate that a linear adaptive response to prevalence provides no additional advantage over a constant adaptive fraction in controlling outbreaks, and a highly superlinear response is required to suppress epidemic spread. For a sigmoidal adaptation, we identify conditions under which oscillations in prevalence can emerge, with peak prevalence exhibiting a non-monotonic dependence on the width of the sigmoidal function, suggesting an optimal parameter range that minimizes epidemic severity. We obtain prevalence, final epidemic size, and peak prevalence as functions of adaptation parameters in all adaptation scenarios considered, providing a comprehensive characterization of the effects of spatial adaptation based on global prevalence information in shaping adaptive epidemic dynamics.
\end{abstract}


\maketitle

\section{Introduction}
\label{sec0}
Mathematical modeling of epidemics plays an important role in understanding and predicting the dynamics of infectious diseases~\cite{Ottar2018,WangStati,Keeling2008,Anderson1985,Rothman2008,schn23,Ashby2021, Fred2001}. When combined with knowledge from diverse disciplines and real-world data, epidemic models help us to make informed decisions to manage pandemics effectively. By integrating various pharmaceutical and non-pharmaceutical interventions, epidemic models provide quantitative framework for suggesting better control measures and mitigating the impact of outbreaks~\cite{wong2020,Markosocial,Funk2009,Buscarino2008, Xinyu2025}.

Compartment models and their several variants have long served as the foundation of epidemic modeling~\cite{Kermack27, Vinc1978, Peng2019, Song2009, Khan2018}. These models divide the population into various compartments based on the state of infection of an individual, the typical compartments being susceptible (S), infected (I), recovered (R), exposed (E), etc. The conventional compartmental models assume a well-mixed population and can be easily extended to consider populations with specific contact structures using network-based approaches~\cite{Turner2020, Pastor2015}. 

A key assumption in conventional compartmental models is that the contact rate between susceptible and infected individuals remains constant over time~\cite{Kermack27}. However, in most situations, as the prevalence of infection increases within a population, the interaction between susceptible (S) and infected (I) individuals may decrease, reflecting a behavioral adaptation effect. An earlier example of an epidemic model accounting such behavioral changes was by Vincenzo et al~\cite{Vinc1978}. The effect of complex social responses on epidemic dynamics has since been explored using models incorporating saturating or non-monotonic interaction functions between the susceptible and the infected population~\cite{Dongmei2007, Ruan2003,Kar2019, Khan2018}. The dynamical behavior of epidemic models with such nonlinear incidence rates has been extensively studied in the past providing a deeper understanding of how human behavior may influence epidemic dynamics (for a review of such models See~\cite{Funk2010}).

A key driver of adaptive behavior is risk perception, which influences precautionary actions such as social distancing, reduced mobility, and mask usage~\cite{Moinet2018, Hehuang2023, Rongcheng2023, DanYang2024, Kar2019}. Risk perception may be based on local information, such as the infection status of immediate neighbors in a contact network, or on global information, such as population-wide prevalence reported by media or public health agencies. Several studies have modeled risk perception as a probabilistic response function influencing individual behavior~\cite{Cui2008, Liu2007,chang2021,Paulo2022, Arthur2021, Vrugt2020, Sharma2019, Maier2020, Nowak2011, Caley2008, Mgosac21, Glaubitz2020, Xinyu2025}. Studies show that a sufficiently strong nonlinear increase in risk perception can lead to epidemic extinction in specific network structures~\cite{Bagnoli2007, Herrera2019}, highlighting the importance of adaptive behavior by individuals in shaping epidemic outcomes.

In addition to behavioral adaptations, the spatial nature of disease transmission has received increasing attention in recent years~\cite{Barthelemy2011,Akhil24, Huang2016, loring2020}. Accounting for this spatial effects is particularly relevant for airborne diseases such as COVID-19, where physical distance between individuals and the spatial contact structure (as opposed to the topological contact structures typically assumed in conventional models) plays a critical role in transmission dynamics~\cite{Gross2020, Peng2019}. In the spatial models of epidemic, agents are assumed to occupy positions in a plane and two agents form a potential disease transmitting link between them only when their physical separation is closer than a particular value. Just as in the conventional framework, one may consider a well-mixed spatial population or a static spatial contact network as representing two opposite extremes of agent mobility. The latter typically make use of the language of Random Geometric Graphs and has garnered increased attention in the recent past~\cite{Saha2023,Spatial2020,loring2020}. Such graphs have a close association with the continuum percolation models in statistical physics and in the context of epidemic spreading, the percolation threshold can act as a determining factor in deciding whether a disease undergoes large-scale spreading in the population or not~\cite{Akhil24,Mello2021}.  

In this work, we integrate the two key aspects discussed above-prevalence-based behavioral adaptation and spatial constraints-to analyze epidemic dynamics in a spatial setting. Specifically, we consider a spatial Susceptible-Infected-Recovered (SIR) model with global incidence-based adaptation by a fraction of agents in the population. The assumed spatial nature of the process naturally lends itself to incorporate various non-pharmaceutical interventions, such as reduction in mobility, adherence to social distancing, and the use of protective measures like masks, all involving a spatial scale~\cite{chang2021,Paulo2022, Arthur2021, Vrugt2020, Sharma2019, Maier2020, Nowak2011, Caley2008, Mgosac21, Glaubitz2020, Xinyu2025}. There are a few recent studies which look into various adaptive mechanisms implemented at both the global and individual levels, examining scenarios where adaptation is informed by prevalence information or by local interactions. Oscillatory dynamics in prevalence have been observed in models where adaptation depends on prevalence and feature a well-defined adaptation threshold for containing the epidemic from spreading~\cite{Jianping2021, Glaubitz2020, Akhil24}.

In this paper, employing a spatial setting, we analyse the impact of three different global-prevalence-based adaption mechanisms employed by the agents on the dynamics of an epidemic. In all cases, we assume that a fraction of the population takes adaptive action based on the global prevalence of the disease. The three scenarios differ in the functional dependence of the adaptive fraction on the global prevalence and cover a wide spectrum of potential behavioral responses. Specifically, the three scenarios correspond to a constant adaptive fraction independent of prevalence, a power-law form and a sigmodial form. In the spatial model, it is assumed that an infected agent has a characteristic transmission radius which is the distance up to which it can potentially pass on the disease. Alternatively, a susceptible agent can catch the disease from any of the infected agents within a distance equal to the characteristic transmission radius. Adaptation is modeled as a reduction in the transmission radius by a fixed factor, which may be implemented by either infected or susceptible individuals, or both. For each of the relationship forms between the adaptive fraction and the global prevalence considered, we treat these three cases separately. In all scenarios, we consider the two extreme limiting cases of spatial mixing, namely, agents getting spatially fully mixed over time and agents remaining spatially static. 

For each adaptation scenario, we analytically determine the critical adaptation thresholds in a spatial mean-field setting and obtain the phase diagram showing the epidemic and non-epidemic phases. For each case, we obtain the key epidemic metrics like prevalence, the final size of the epidemic, and peak prevalence as a function of the adaptation parameters. We do this analytically numerically in the spatially well-mixed regime and using simulations in the spatially static regime. In addition, with static agents, we obtain useful bounds for the adaptation thresholds by drawing parallels to continumm disc percolation problems with a distribution of radii~\cite{Meeks2017a, Meeks2017b,Mertens2012,Phani1984}.

In the case of a constant adaptive fraction, our results demonstrate that for a given set of disease parameters and population density, there is a minimum level of adaptation by the agents required to contain the epidemic. If adaptation is below this level, then the epidemic will spread in the spatial population even if all agents take adaptive action.  In the case of a power-law form of adaptive fraction, we show that an extremely superlinear adaptive response by the agents towards global prevalence will be required to effectively contain the epidemic. In this case, we also analyze the effect of imperfect prevalence information on agent adaptation and its influence on the critical parameters. Finally, for the sigmoidal type adaptation by the agents, our results show that, it is effective in preventing the epidemic only within a narrow range of parameters. In this case, we identify a regime in which oscillations in prevalence emerge when the width of the sigmoidal response is sufficiently small. Moreover, we find that the peak prevalence exhibits a non-monotonic dependence on the width of the sigmoid, revealing the existence of an optimal adaptation range that minimizes epidemic severity.

The remainder of this paper is organized as follows. In Sec.~\ref{sec1}, we define a spatial, discrete-time SIR epidemic model with prevalence-based adaptation by the agents and define the relevant parameters.  In the following  sections, Sec.~\ref{sec2} - Sec.~\ref{sec4}, we precisely define and analyze the three adaptation scenarios - constant, power-law type and sigmoidal type - in detail. We obtain the critical parameters in the spatially well-mixed case, obtain bounds for them in the spatially static case, and study the detailed evolution of the epidemic. We obtain numerical and simulation results for prevalence, final epidemic size, and the peak prevalence, comparing the different adaptation models. We conclude in Sec.~\ref{sec5},  summarizing our findings and discussing potential extensions. 

\section{Spatial SIR model with agent adaptation}
\label{sec1}
We consider a Susceptible-Infected-Recovered (SIR) epidemic model in which agents are distributed randomly within a 2D plane with spatial density $\rho$. We assume that each agent in the system is surrounded by a circular region of a certain radius centered on its present position termed as the Characteristic Range (CR)~\cite{Akhil24,Huang2016}. Consider time as discrete, and if the focal agent is an infected one, it can transmit the disease to a susceptible one present within the region with a probability $\beta$ - the transmission probability - at each instant. Alternatively, if the focal agent is a susceptible one, it can catch the disease from an infected agent in its surrounding circular region with probability $\beta$ at each instant. An infected agent recovers and becomes immune to infection at each instant with a probability $\gamma$ - the recovery probability.

Behavioral adaptation by the agents is incorporated in the model by assuming  that a fraction of the susceptible and/or infected agents can actively reduce the radius of the CR surrounding their position by a certain amount. Infected agents may achieve this through interventions such as mask-wearing, social distancing, or avoiding crowded areas, thereby decreasing their effective transmission range. Similarly, susceptible agents may reduce their likelihood of infection by engaging in protective behaviors that limit their exposure radius.  We examine three adaptation scenarios that capture a broad range of potential behavioral responses.  Specifically, we consider the three cases in which the fraction of agents who engages in adaptation is {\bf A)} constant, {\bf B)} depends on the global prevalence of infection in a power-law manner, and {\bf C)} depends on the global prevalence in a sigmoid-law manner. 

\begin{figure}[!h]
    \centering
    \includegraphics[width=8cm]{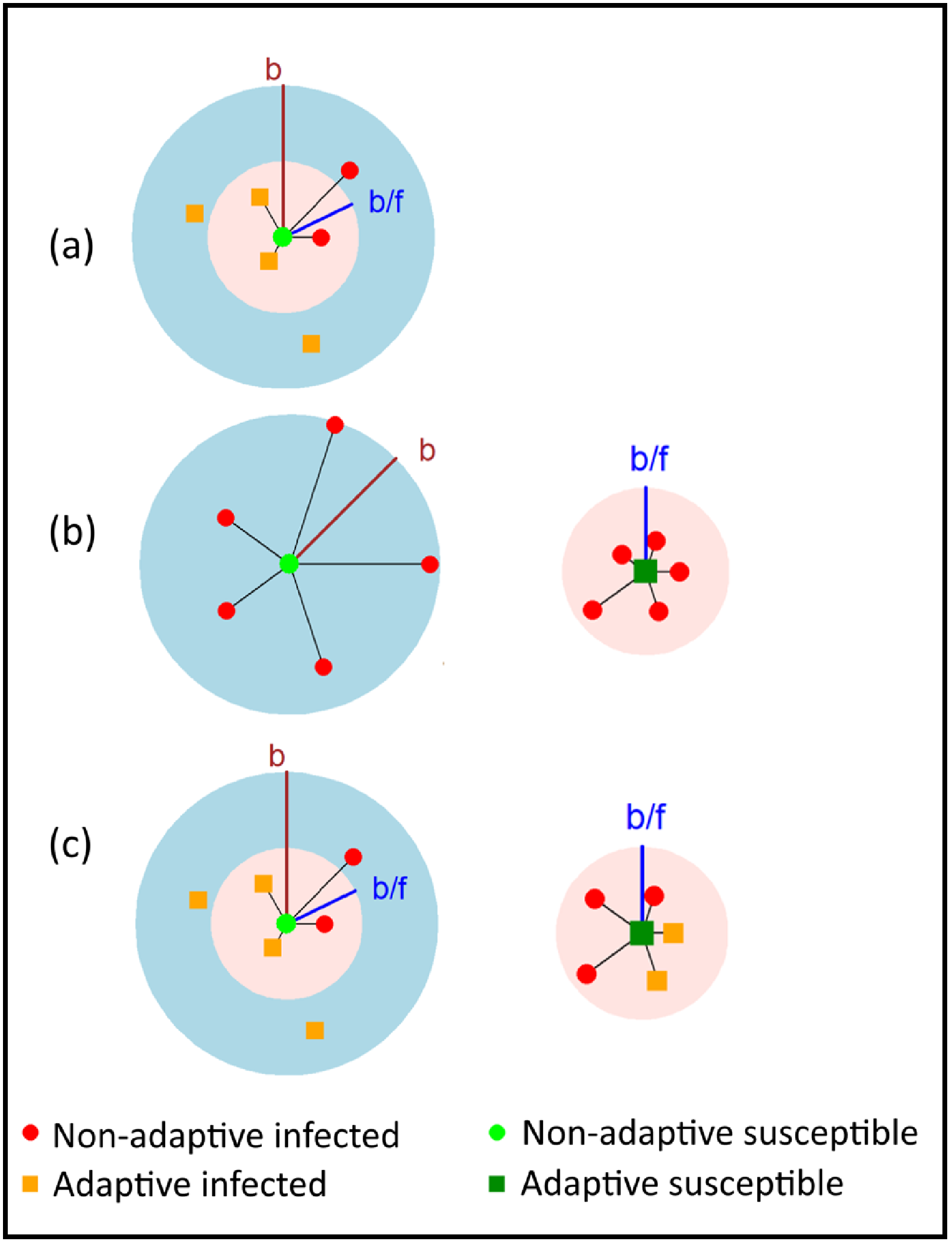}
    \caption{Schematic of AI, AS, and AIS scenarios. The presence of a line between two nodes indicates the potential for disease transmission between them.\\
(a) For the AI case, a fraction of the infected agents reduces their  Characteristic Range (CR) of disease transmission by a factor $f$ termed as the adaptation factor. Thus, each  susceptible agent can acquire infection from all the infected agents within a distance
$b/f$ from its location but only from the non-adaptive infected agents with the higher CR of $b$ located within the
annular region between distances $b$ and $\frac{b}{f}$. \\
(b) For the AS case, a fraction of the susceptible agents take adaptive action and reduce their CR by a factor $f$. Thus, a non-adaptive susceptible agent can acquire the infection from all the infected agents within a distance $b$ from its location, whereas an adaptive susceptible agent can acquire the infection only from infected agents within a distance $b/f$. 
(c) For the AIS case, a fraction of both the susceptible and infected agents take adaptive action and reduce their CR by a factor $f$. Therefore, a non-adaptive susceptible agent can acquire infection from all the infected agents within a distance
$b/f$ from its location but only from the non-adaptive infected agents with the higher CR of $b$ located within the
annular region between distances $b$ and $\frac{b}{f}$. An adaptive susceptible agent can acquire the infection from all infected agents within a distance $b/f$ from its location.}
    \label{schematic}
\end{figure}

In general, the CR can have a distribution as considered in ~\cite{Huang2016} that could also be varying with time. Here, for simplicity, we assume that all non-adaptive agents have a uniform CR of $b$, and all those who adapt reduce their CR by a factor $f$ where $f>1$~\cite{Akhil24} which we refer to as the adaptation factor. For the cases A, B, and C described above, we analyse three possible scenarios namely, i) Adaptive Infected (AI) - Only infected agents reduce their CR, ii) Adaptive Susceptible (AS) - Only susceptible agents reduce their CR, and iii) Adaptive Infected \& Susceptible (AIS) - Both infected and susceptible agents reduce their CR. The three scenarios are depicted in Fig.~\ref{schematic}. For adaptation scenario B, we also investigate the effect of adaptation by the agents under conditions of imperfect global prevalence information.

Agents redistribute themselves within the two-dimensional plane at each time step. While various models for agent mobility can be considered, to delineate the effect of agent adaptation, we focus on the following two limiting cases. In the first case, agents redistribute uniformly randomly in each time step making it possible to analyze the problem using a spatial mean-field type analysis.  In the second case, agents remain fixed in their initial locations in space over time, leading to an epidemic process that unfolds on a random geometric graph, where two nodes are connected if their spatial distance is less than the CR~\cite{Barthelemy2011,Akhil24, Huang2016, Saha2023}.

 \section{Constant adaptation}
 \label{sec2}
In this scenario, at each instant, a {\it constant } fraction of the infected (I) and/or susceptible (S) agents adapt by reducing their CR $b$ by a factor $f$. We will consider the cases AI, AS and AIS in order below.
\subsubsection{Adaptive Infected (AI)}
\label{constant_ai}
Assume that a fraction $k_{I}$ of the infected population engages in adaptive behavior and reduces their CR to $\frac{b}{f}$ ($f> 1$), while the remaining $(1 - k_{I})$ fraction retains the original CR $b$. Consider time $t$ as discrete and at each instant, agents are randomly repositioned in the 2D plane, thus creating a homogeneously mixed population over time. Since there are infected agents with two different CRs, a focal susceptible agent can acquire infection from all the infected agents within a distance $\frac{b}{f}$ from its location but only from the non-adaptive infected agents with the higher CR of $b$ located within the annular region between $\frac{b}{f}$ and $b$ from its location (See  Fig.~\ref{schematic}\textbf{\textit{a}}).

From Fig.~\ref{schematic}\textbf{\textit{a}}, it is clear that on an average there will be $\rho \pi (b/f)^2 i(t) $ infected agents in the inner circle of a focal susceptible agent  and $\rho \pi (b^2 - (b/f)^2)i(t)(1 - k_I)$ infected agents in the annular region . Noting that $(1-\beta)^{N_{I}}$ is the probability that a susceptible agent does not get infection from any of the $N_I$ agents located in its \textquoteleft neighborhood\textquoteright\; from whom it can potentially get infection in a time step, we can write the evolution equation for the fraction of agents in the S, I and R compartments as,

\begin{equation}
\begin{split}
    s(t+1) & = s(t)  \\
    & - s(t) \left(1 - \left(1 - \beta\right)^{\frac{\rho \pi b^{2}}{f^{2}}\left(1+(f^{2}-1)(1-k_{I})\right)i(t)}\right)\label{eqn1I}
\end{split}
\end{equation}
\begin{equation}
\begin{split}
    i(t+1) &= i(t) \\
    & + s(t) \left(1 - \left(1 - \beta\right)^{\frac{\rho \pi b^{2}}{f^{2}}\left(1+(f^{2}-1)(1-k_{I})\right)i(t)}\right)\label{eqn2I} \\
    & - \gamma i(t)
\end{split}
\end{equation}
\begin{equation}
\begin{split}
    r(t+1) = r(t)  + \gamma i(t)\label{eqn3I}
\end{split}
\end{equation}
We will focus on the evolution of the infected fraction given by Eq.~\ref{eqn2I}, and determine the critical value of the adaptive fraction $k_I$ for a given adaptation factor $f$, for which a small outbreak of the disease will not spread in the population. In Eq.~\ref{eqn2I}, neglecting terms of higher orders in $\beta$ and assuming an early phase of the epidemic during which $r(t)$ is very small so that $s(t) \approx 1 - i(t)$, we can write,
 \begin{equation}
 \begin{split}
    i(t+1) & \approx i(t)+ \\
    & \left(1-i(t)\right) \rho \pi \beta \frac{b^{2}}{f^{2}}  i(t)\left(1+(f^{2}-1)(1-k_{I})\right) - \gamma i(t) \label{eqn5}
    \end{split}
\end{equation}   
Keeping only terms linear in $i(t)$ which is justified in the initial phase of an epidemic during which $i(t) << 1$, Eq.~\ref{eqn5} becomes, 
\begin{equation}
    \frac{i(t+1)}{i(t)} \approx 1 + \frac{\rho \pi \beta b^{2}}{f^{2}}\left(1+(f^{2}-1)(1-k_{I})\right) - \gamma \label{eqn7}
\end{equation}
For the disease to not grow in the population, we require  $
 i(t+1)/i(t)<1$. Using this condition, the critical fraction of the infected population who must adapt (reduce their CR by a factor $f$) for the epidemic to not grow in the population is given by,
\begin{equation}
    k^{crit}_{I } =1 - \frac{\gamma f^{2} - \rho \pi \beta b^{2}}{\rho \pi \beta b^{2}(f^{2} - 1)}
    \label{eqn10}
\end{equation}
Thus for a spatially well-mixed population with given values of $\rho$, $\beta$, $\gamma$, and $b$, the above equation gives the minimum fraction of infected agents who must reduce their CR by a factor of $f$ in order to contain the epidemic. The combination $\rho \pi \beta b^2$ will occur often that we will denote it by the symbol $\Delta$. We can infer that $\Delta$ represents the average number of disease-causing connections a non-adaptive infected agent has at each instant.  We can rewrite Eq.~\ref{eqn10} as,
\begin{equation}
    k^{crit}_{I } =1 - \frac{\gamma f^{2} - \Delta}{\Delta(f^{2} - 1)}
    \label{kIcrit}
\end{equation}

\begin{figure}[h]
    \centering
    \includegraphics[width=8.5cm]{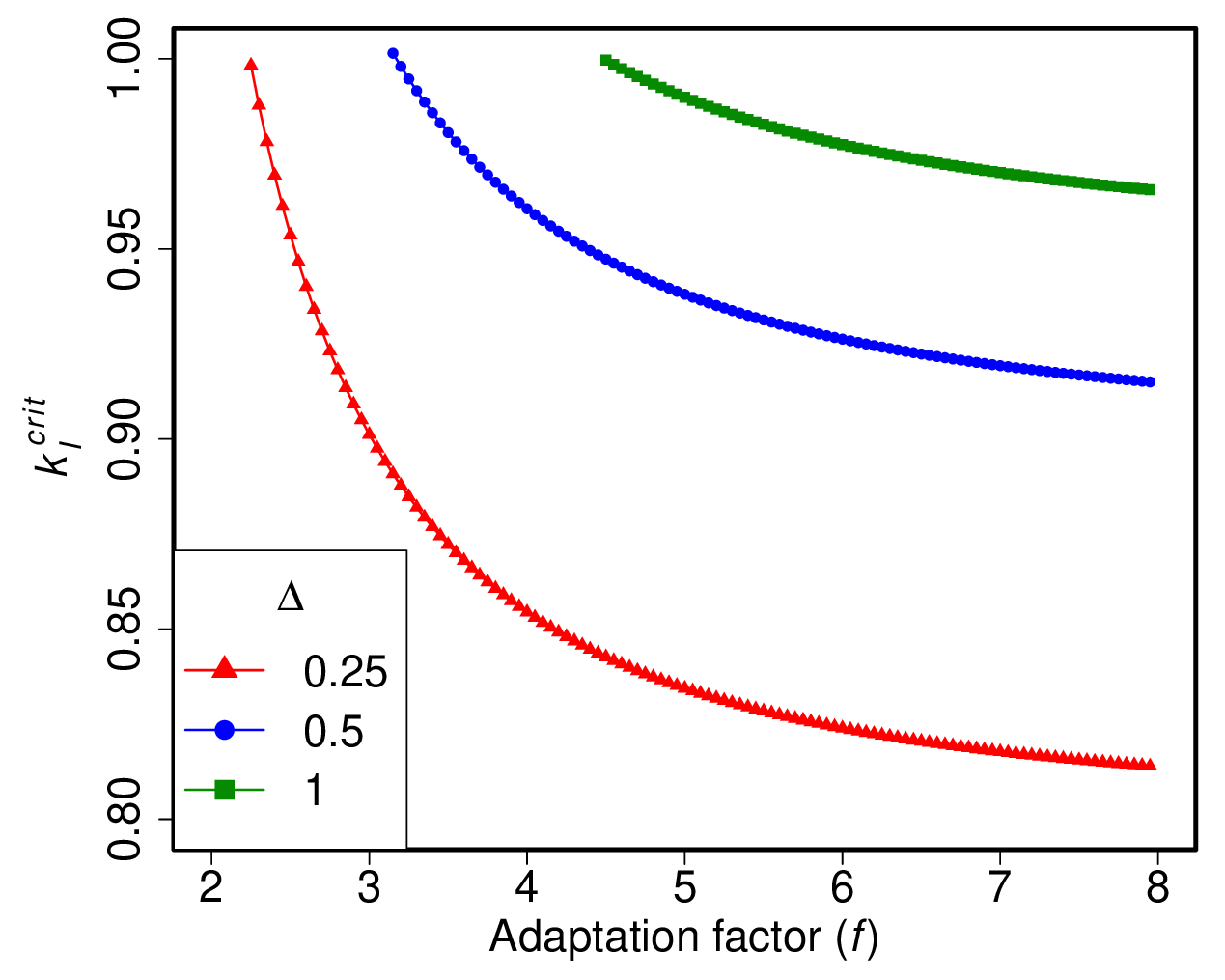}
    \caption{Variation of the critical fraction of infected agents $k_{I}$ with adaptation factor $(f)$ for various values of the parameter $\Delta = \rho\pi\beta b^2$.} 
    \label{fig:kicrit}
\end{figure}

Fig.~\ref{fig:kicrit} shows the variation of $k^{crit}_{I}$ with $f$ for different values of $\Delta$. For each value of $\Delta$, the area below the curve corresponds to the epidemic phase where the infection spreads in the population, whereas the area above represents the non-epidemic phase.  Note that for a given set of parameters, there is a minimum value of $f$ say $f_{min}$ below which adaptation becomes ineffective in containing the epidemic. We can obtain $f_{min}$ from Eq.~\ref{kIcrit} by determining the value of $f$ below which the $k^{crit}_{I}$ becomes greater than one so that even if all infected agents adapt, the disease will still spread in the population. This happens when $\gamma f^2 < \Delta$ in Eq.~\ref{kIcrit}. Therefore, 
\begin{equation}
  f_{min} = \sqrt{\Delta/\gamma}\label{fmin_ai}
\end{equation}

Also, note that as $f \rightarrow \infty$, $k^{crit}_{I} \rightarrow 1 - \gamma/\Delta$ in Eq.~\ref{kIcrit}. This corresponds to the case where adaptive infected agents reduces their CR to zero, thereby effectively removing themselves from the population. Here, $\Delta/\gamma$ is analogous to the basic reproduction number $R_0$ in the usual setting~\cite{Ottar2018, Rothman2008, Aparicio2007}. The usual condition for containing a disease from spreading is to vaccinate (and thereby remove) at least a fraction of the population $(1 - 1/R_0)$ ~\cite{Ashby2021}. Likewise in the spatial setting described, the minimum fraction of infected agents who should take adaptive action (reduce their CR by the adaptation factor $f$) in order to contain the disease from spreading is given by Eq.~\ref{kIcrit}.  Here, spatial effects are incorporated via density of agents, CR, and f. While vaccination is assumed to completely remove an infected agent from the population, taking spatial adaptive measures like reducing mobility or taking protective measures like wearing a mask is assumed to reduce their CR by a fixed factor. 

We can calculate how the conventional expressions for the peak prevalence and the final size of the epidemic are modified due to the spatial adaptation. From the linearized version of Eq.~\ref{eqn2I}, the fraction of susceptibles when the prevalence reaches its maximum is given by,
\begin{equation}
    s_{M} = \frac{\gamma}{\eta \Delta}
\end{equation}
Where $\eta$ is given by,
\begin{equation}
    \eta_{AI}  = \frac{1 + (f^{2} - 1 )(1 - k)}{f^{2}}\label{etaAIk}
\end{equation}
On integrating the expression for $\frac{i(t+1) - i(t) }{s(t+1) - s(t)}$ treating $t$ as continuous, and using $s_{M}$ gives the expression for the maximum prevalence,

\begin{equation}
    i_{max} = -\frac{\gamma}{\eta\Delta} + \frac{\gamma}{\eta\Delta}\ln(\frac{\gamma}{\eta\Delta}) + s_{0} + i_{0} - \frac{\gamma}{\eta\Delta}\ln(s_{0})\label{Imaxeqn}
\end{equation}
where $s_0$ and $i_0$ are the initial values of $s(t)$ and $i(t)$ respectively. 

A self consistent equation for the final epidemic size $r_{\infty}$ can be calculated by considering the ratio $\frac{s(t+1) - s(t) }{r(t+1) - r(t)}$, which gives,
\begin{equation}
   r_{\infty} = 1 -  s_{0}\exp {-\dfrac{\eta\Delta r_{\infty}}{\gamma}}
   \label{etaAI_r}
\end{equation}

When the agents are spatially static, we expect a lower threshold value for the adaptive fraction compared to the well-mixed scenario. Hence, the threshold given by Eq.~\ref{kIcrit} for the well-mixed case can act as an upper bound for the  corresponding static one. A lower bound for the static case can be obtained by considering the spatial system as made up of overlapping discs of two radii corresponding to the adapted and non-adapted infected fractions. Since there is no mixing of agents over time in the static case, the spreading of the disease is influenced by the CR of infected  agents alone. For the static case with AI agents, at any instant, assume that the system is composed of agents of two different CRs, namely, a fraction $\rho k_I$ having radius $b$ and a fraction $\rho (1-k_I)$ having radius $b/f$. We can view this as a continuum percolation system with overlapping discs of two different radii. Then, a necessary condition for the disease to spread in the spatially static population is that these discs form a giant percolating cluster. If $A_c$ denotes the percolation threshold (in terms of areal density) for such an overlapping disc system of two possible radii, then the condition for containing the epidemic is that the sum of the areal densities of the two types of discs should be less than the percolation threshold $A_c$. I.e,  

\begin{equation}
    \rho k_I \pi (b/f)^2 + \rho (1-k_I) \pi b^2 < A_c
    \label{perc2d}
\end{equation}
Therefore, a lower bound for the threshold value of adaptive fraction say $ k_{I_-}^{crit}$ is given by,
\begin{equation}
    k_{I_-}^{crit} = \dfrac{\rho \pi b^2 - A_c}{\rho \pi b^2(1 - 1/f^2)}\label{k_perc}
\end{equation}
We can use $A_c \approx 2.256$ which is the approximate maximum possible value of the areal density for the two-sized disc system respecting the inequality in Eq.~\ref{perc2d}~\cite{Phani1984}. 

\subsubsection{Adaptive Susceptible (AS)}
In this case, a fraction $k_{S}$ of the susceptible population takes adaptive action such that their CR is reduced by the adaptation factor $f$. An adapted agent can catch the disease only from infected agents within a radius of $\frac{b}{f}$. The remaining $(1-k_{S})$ fraction of the susceptible agents can get the disease from infected agents within a radius of $b$ (See Fig.~\ref{schematic}\textbf{\textit{b}}). Similar to the AI case, we can write the evolution equation for the fraction of agents in the infected compartment, 
\begin{equation}
\begin{split}
    i(t+1) &= i(t)\\
    & + s(t)(1-k_{S}) \left(1 - \left(1 - \beta\right)^{\rho \pi b^{2}i(t)}\right)\\
    & + s(t)k_{S}\left(1 - \left(1 - \beta\right)^{\rho \pi\frac{b^{2}}{f^{2}}i(t)}\right) - \gamma i(t)\label{eqnS2}
\end{split}
\end{equation}

Though Eq.~\ref{eqnS2} and Eq.~\ref{eqn2I} are different, it is easy to show that for small values of $\beta$ and in the early phase of an epidemic where $i(t)$ is small and $r(t)$ is negligible, Eq.~\ref{eqnS2} reduces to the same form as Eq.~\ref{eqn5} with $k_S$ in place of $k_I$. This implies that the critical fraction of the susceptible agents say $K_{S}^{crit}$ is equal to $K_I^{crit}$ given by Eq.~\ref{eqn10} and that $f_{min}$ is given by Eq.~\ref{fmin_ai}. Therefore, although the epidemic takes a different trajectory for the AS case compared to the AI case (for instance, see the prevalence curves for the two cases in Fig.~\ref{fig:AIASASIpre}), their critical values are the same. Also, for the AS case with static agents, we can make an argument similar to the AI case which gives the same bound as in Eq.~\ref{k_perc}. Peak prevalence  and final epidemic size can be calculated using Eq.~\ref{Imaxeqn} and Eq.~\ref{etaAI_r} where, 
\begin{equation}
 \eta =  1 - k + \frac{k}{f^{2}}\label{etaAS}   
\end{equation}   
\subsubsection{Adaptive infected and susceptible (AIS)}
\label{ais_constant}
Here both infected and susceptible agents are assumed to take adaptive actions. Assume that $k_{S}$ fraction of the susceptible population adapts and their CR is $\frac{b}{f}$  whereas $(1-k_{S})$ fraction of the agents does not adapt and their CR is $b$. Also, $k_{I}$ fraction of the infected population takes adaptive action (CR $b/f$) and $(1-k_{I})$ fraction of the infected population has a CR of $b$. It is conceivable to have different values of $f$ for the infected and the susceptible agents, but for simplicity, we will assume that it is the same for the two groups (See Fig.~\ref{schematic}\textbf{\textit{c}}). In the AIS case, the equation for the infected compartment becomes,

\begin{equation}
\begin{split}
\small
    i(t+1) &= i(t)\\
    &+ s(t)(1-k_{S})\left(1 - \left(1 - \beta\right)^{\rho \pi\frac{b^{2}}{f^{2}}i(t)+\rho \pi (b^{2}-\frac{b^{2}}{f^{2}})(1-k_{I})i(t)}\right)\\
       &+ s(t)k_{S} \left(1 - \left(1 - \beta\right)^{\rho \pi \frac{b^{2}}{f^{2}}i(t)}\right) - \gamma i(t) \label{eqnASI}
\end{split}
\end{equation}

To simplify further, we will assume that the fractions of the infected and susceptible agents who adapt are the same $k_S = k_I = k_{SI}$. Note that these fractions are defined with respect to the number of agents in the different compartments and not with respect to the total population. During the initial phase of an epidemic, there will be a lot more agents in the susceptible compartment compared to the infected one. From Eq.~\ref{eqnASI}, we can obtain the critical fraction $k_{SI}^{crit}$ as,
\begin{equation}
     k^{crit}_{SI}  = 1 - \sqrt{ \frac{\gamma f^{2} - \Delta}{\Delta(f^{2}- 1)}} \label{KSIcrit} 
\end{equation}
The lower limit of $f$ below which $k_{SI}^{crit}$ is greater than one is still given by Eq.~\ref{fmin_ai}. Comparing Eqs.~\ref{kIcrit} and \ref{KSIcrit}, we can see that for allowed values of $f$, the two critical values are related by the expression,
\begin{equation}
k^{crit}_{SI} = 1 - \sqrt{1+k^{crit}_I}
\label{krelation}
\end{equation}
Peak prevalence  and final epidemic size can be obtained using  Eq.~\ref{Imaxeqn} and Eq.~\ref{etaAI_r} where, 
\begin{equation}
   \eta =  \frac{1+ (1-k)^{2}(f^{2} - 1)}{f^{2}}\label{etaAIS}
\end{equation}
Finally, for the AIS case with static agents, we can make an argument similar to the AI case which gives the same bound as in Eq.~\ref{k_perc}. 

\begin{figure}
    \centering
    \includegraphics[width=1\linewidth]{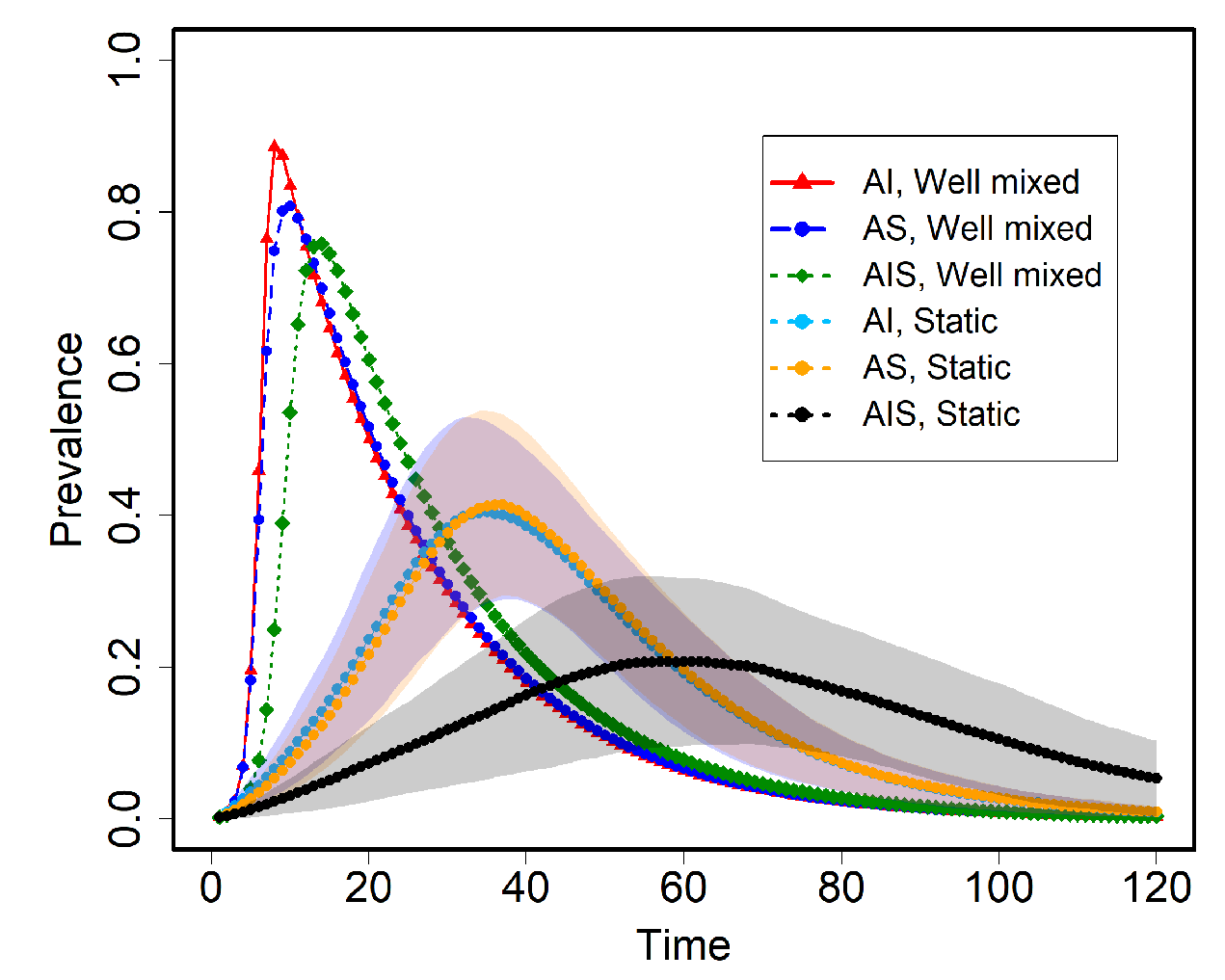}
    \caption{Plots showing typical variation of prevalence with time when a constant fraction $k$ of the agents take adaptive action in the AI, AS, and AIS cases. For each case, prevalence in a spatially well-mixed scenario (obtained numerically from Eq.~\ref{eqn2I}, Eq.~\ref{eqnS2} and Eq.~\ref{eqnASI})  is compared with that of spatially static agents (obtained by monte carlo simulations). For the static case, the solid curves represent the average values obtained in $10^3$ trials and the shadows represent the spread.}
    \label{fig:AIASASIpre}
\end{figure}

Having analysed the critical thresholds with constant adaptation for the AI, AS, and AIS models in the spatially well-mixed and static regimes, we now look into their dynamic behaviour in detail and compare the results. For the well-mixed scenario, the prevalence curves for the AI, AS, and AIS cases are obtained by solving the evolution equations Eqs.~\ref{eqn2I}, \ref{eqnS2} and~\ref{eqnASI} numerically and for the static case, the results are obtained by monte carlo simulations of the model on random geometric graphs. For obtaining numerical results in either case, we use the values $\rho = 400, \beta = 0.5, \gamma = 0.05, b = 0.07, i(0) = 1/400$ and $ f = 25$ unless otherwise specified. These choices correspond to $\Delta \approx 3.08$. Fig.~\ref{fig:AIASASIpre} shows a comparison of the evolution of prevalence with time for the AI, AS, and AIS cases under constant adaptation for the fully mixed and static scenarios for a specific set of parameter values.  We can see the effects of both the mobility of the agents (well-mixed Vs static) and the type of adaptation (AI Vs AS Vs AIS) on the time evolution of the epidemic in the figure. 

\begin{figure}
    \centering
    \includegraphics[width=1\linewidth]{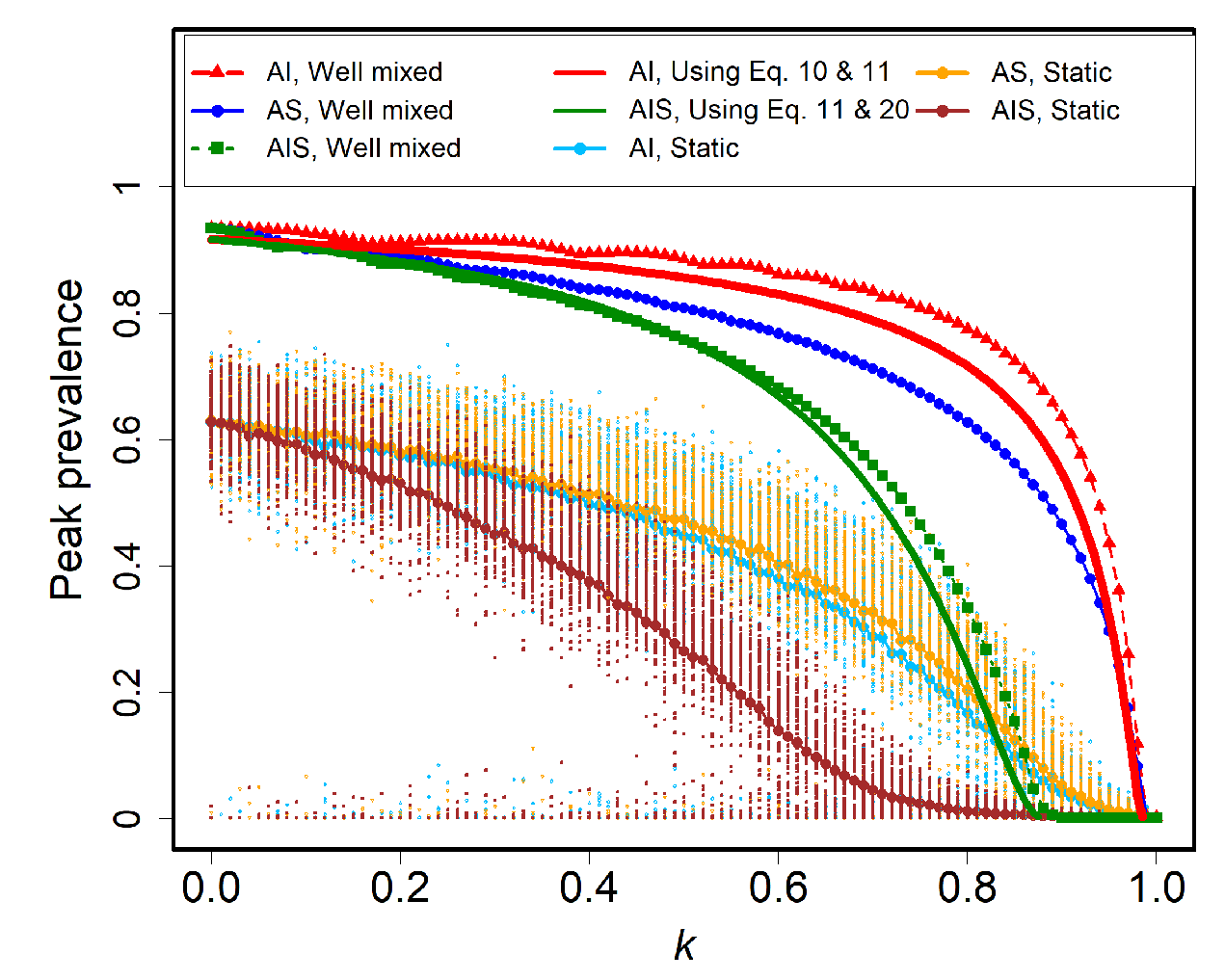}
    \caption{Variation of peak prevalence with adaptive fraction $k$ for  the AI, AS, and AIS cases. Peak prevalence for the spatially well-mixed case and static case are shown for each scenario. 
    For the AI and AS cases, the transition point for the well-mixed case obtained from Eq.~\ref{kIcrit} is 
    $k^{crit}_{I} = k^{crit}_{S} \approx 0.985$, and for AIS case, it is 
    $k^{crit}_{IS} \approx 0.8787$ (from Eq.~\ref{KSIcrit}). The lower bound for the three static cases obtained from Eq.~\ref{k_perc} is $k^{crit}_{I-} \approx 0.63$.}
    \label{fig:AIASASIpp}
\end{figure}

\begin{figure}
    \centering
    \includegraphics[width=1\linewidth]{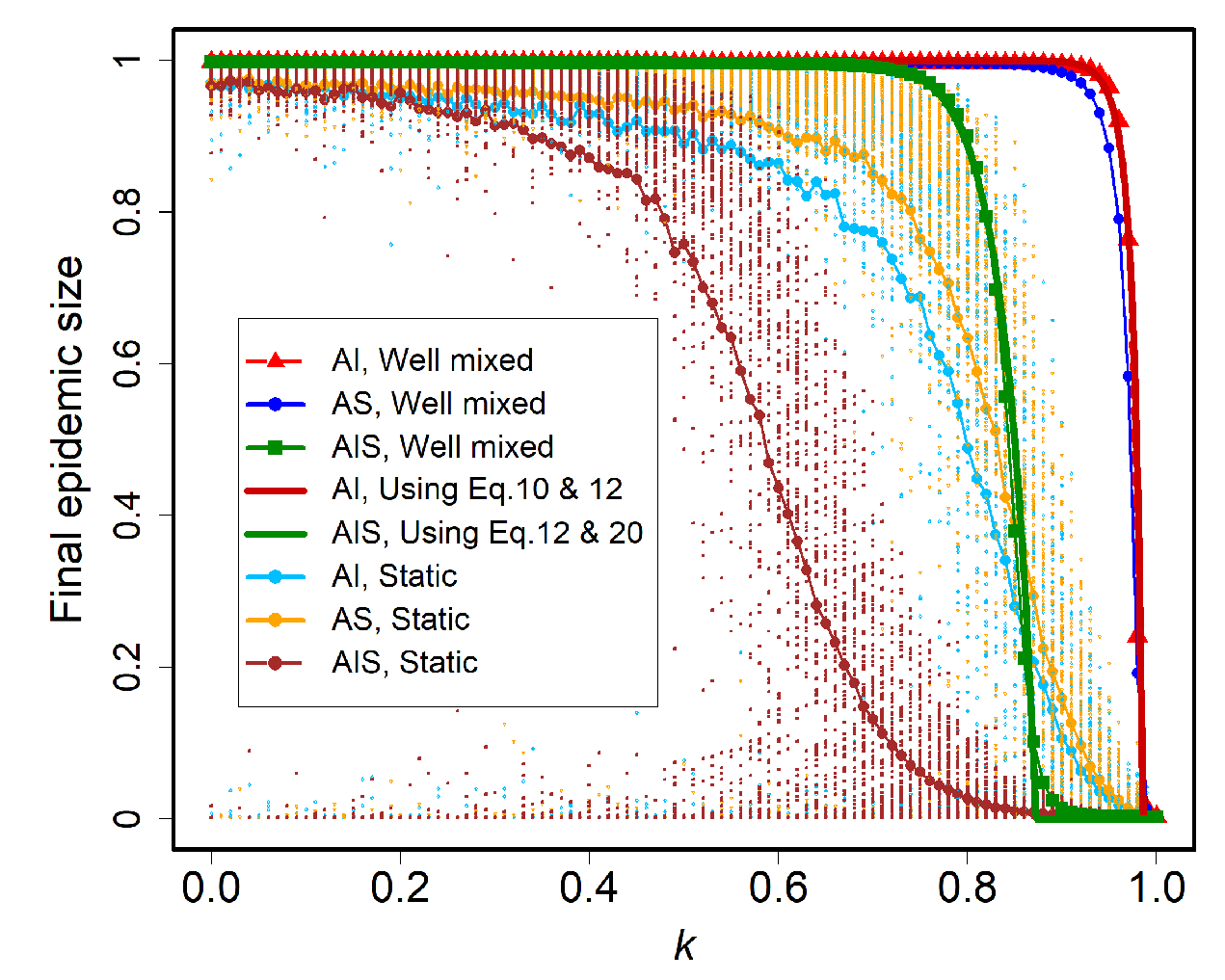}
    \caption{Variation of the final epidemic size with adaptive fraction $k$ for the AI, AS, AIS scenarios.}
    \label{fig:AIASASIfe2}
\end{figure}

Fig.~\ref{fig:AIASASIpp} and Fig.~\ref{fig:AIASASIfe2} respectively show the variation of peak prevalence and final size of the epidemic with adapted fraction $k$ depicting the threshold nature of the epidemic process  as a function of $k$. Comparing the different cases, we can see from Fig.~\ref{fig:AIASASIpp} that AIS model has the lowest threshold and lower peak prevalence levels for all values of $k$. We can also see that although the threshold values of $k$ are the same for AI and AS models, the peak prevalence is lower for the latter indicating that susceptibles reducing their CR is more effective in bringing down the peak infection levels. In Figs.~\ref{fig:AIASASIpp} and~\ref{fig:AIASASIfe2}, we also compare the peak prevalence and final epidemic size with that obtained from Eq.~\ref{Imaxeqn} and Eq.~\ref{etaAI_r} with appropriate values of $\eta$.  From the figures, we can infer that the response of the peak prevalence and final epidemic size to a change in $k$ just below the threshold is more pronounced and highly non-linear for the well-mixed case compared to the static one. We can also see that the threshold values for the well-mixed case obtained from Eqs.~\ref{kIcrit} and \ref{KSIcrit} are in excellent agreement with the numerical solutions. Also, we note that, in many trials in the static case, the epidemic will die down very early simply due to the stochastic nature of the process, resulting in a range of peak prevalence and final epidemic size values for a given $k$ as depicted in Figs.~\ref{fig:AIASASIpp} and~\ref{fig:AIASASIfe2}.

\section{Global information-based adaptation: Power-law}
\label{sec3}
\begin{figure}[h]
    \centering
    \includegraphics[width=1\linewidth]{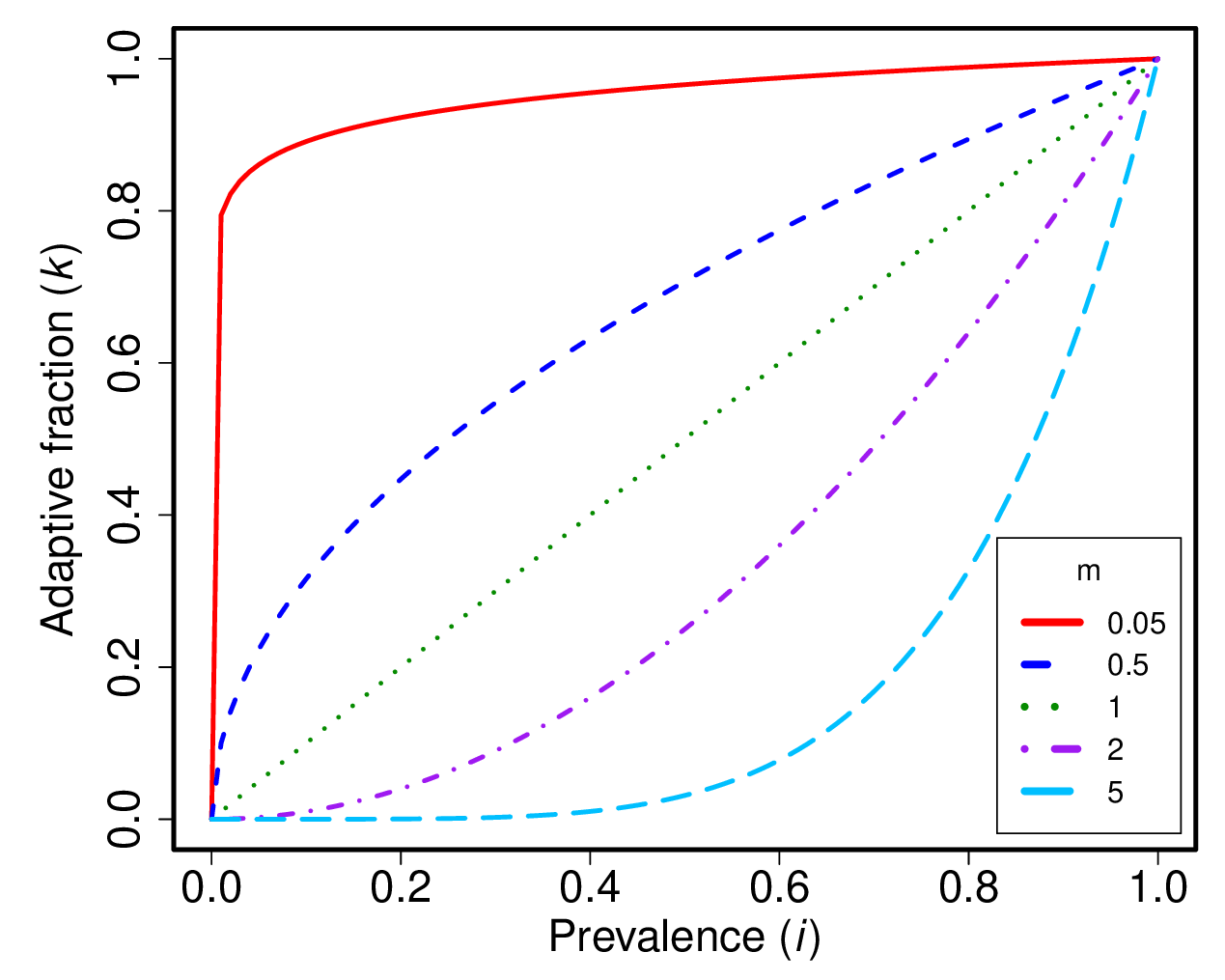}
    \caption{Variation of adaptive fraction $k$ with global prevalence $i$ for different values of $m$ in the power-law model of adaptation where $k = i^m$. $m<1(>1)$ indicates a rapid (slow) initial reaction of the agents to the prevalence information.}
    \label{fig:powerlawplot}
\end{figure}

In this section, we consider the case where the fraction of agents who adapt is not constant during the course of the epidemic but instead depends dynamically on the global prevalence level in a power-law manner.  Specifically, we assume that the fraction who adapts varies as $i(t)^m$ where $m>0$. Thus, $m>1$ corresponds to a sublinear response in the early stages of the outbreak (when $i(t) << 1$) where the agents react rather slowly to the rising prevalence, while $m<1$ corresponds to a superlinear initial response, where individuals begin adapting more rapidly even when prevalence is low (See Fig.~\ref{fig:powerlawplot}). Thus, the parameter $m$ serves as a quantitative measure of the population's fear response to the epidemic. As in the previous section, we analyze three distinct adaptation schemes: i) Only infected agents adapt (AI) ii) Only susceptible agents adapt (AS) and iii) Both types of agents adapt (AIS).

\subsubsection{Adaptive Infected (AI)}

Here, the fraction of infected agents $k_{I}$ following adaptation is assumed to be,
\begin{equation}
    k_{I}(t) = i(t)^{m}
\end{equation}
Using this form of $k_I$ in Eq.~\ref{eqn5} along with $\Delta = \rho \pi \beta b^2$, we get
 \begin{equation}
 \begin{split}
    i(t+1) & \approx i(t)+ \\
    & s(t) \frac{\Delta}{f^2} i(t)\left(1 +\left(f^2-1\right)(1-i(t)^{m})\right) \\
    & - \gamma i(t) \label{eqnmi}
    \end{split}
\end{equation}   
Note that for $m=1$, keeping only linear terms in $i(t)$ in the above equation leads to the same critical fraction given by Eq.~\ref{kIcrit} in the case of constant adaptation discussed in Sec.~\ref{constant_ai}. Thus a linear response by the agents to the global prevalence information is only as good as a constant response in containing the epidemic. Hence, we will focus on super linear initial response where values of $m$ lie between $0$ and $1$. Assuming that $s(t) \approx (1-i(t))$ at the beginning of the epidemic, and neglecting all terms involving second and higher powers of $i(t)$, Eq.~\ref{eqnmi} becomes,
 \begin{equation}
 \begin{split}
    i(t+1) & \approx i(t)+ \\
    & \Delta \left(i(t)-i(t)^{m+1}+\frac{i(t)^{m+1}}{f^{2}}\right)- \gamma i(t) \label{eqnmAI}
    \end{split}
\end{equation}   
 
\begin{figure}
    \centering
    \includegraphics[width=1\linewidth]{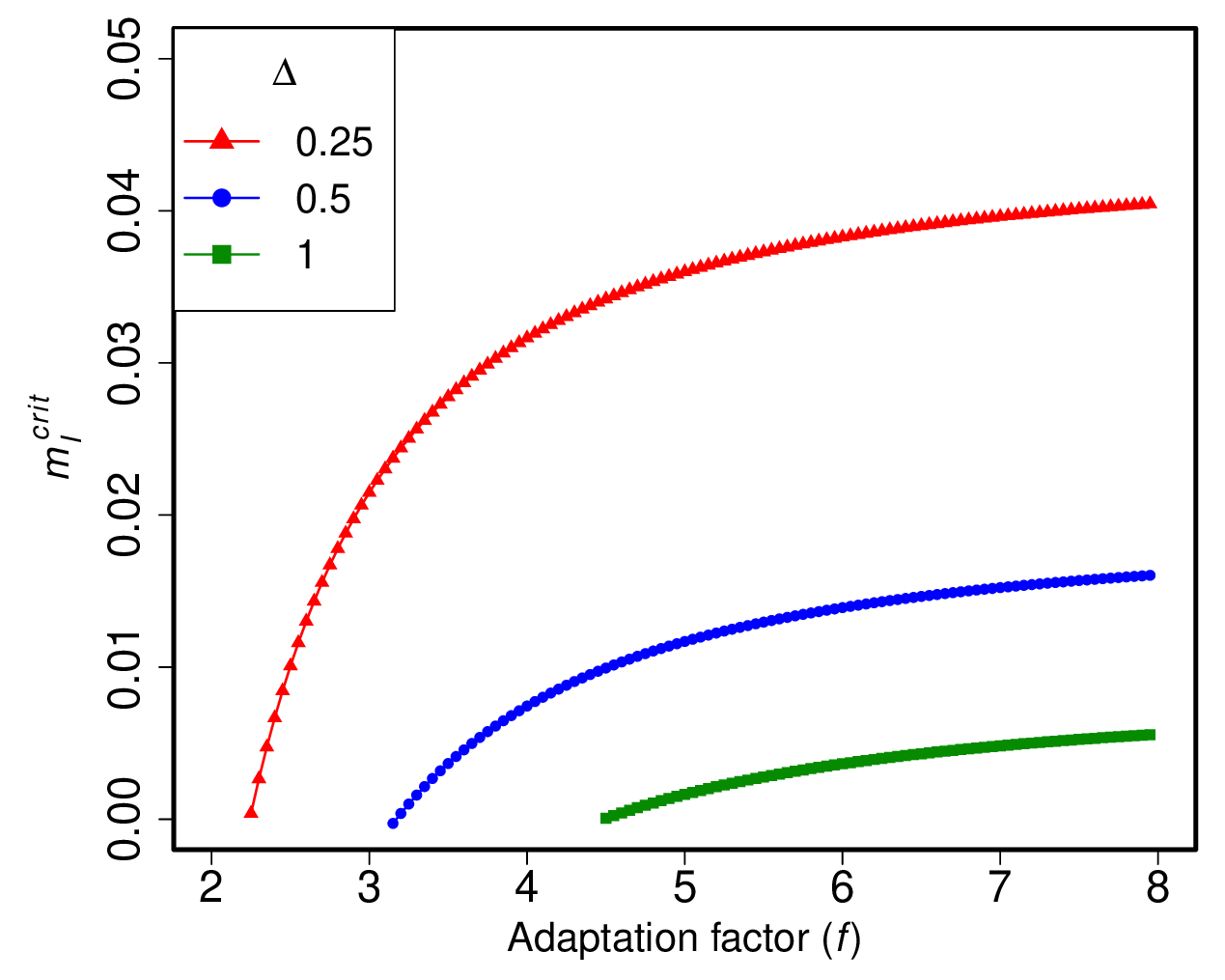}
    \caption{Variation of the critical value $m^{crit}_{I}$ with adaptation factor $f$ for different values of $\Delta$.}
    \label{fig:micrit}
\end{figure}

Because $m<1$, we can expand the term $i(t)^{m+1} \equiv  i(t) \exp \left(\ln (i(t)^m)\right)$ and retain only the term linear in $m$. I.e,
\begin{equation}
    i(t)^{m+1} \approx i(t)\left( 1 + m\ln{i(t)}\right)\label{eqnaprox}
\end{equation}
Using this in Eq.~\ref{eqnmAI}, we get
\begin{equation}
\begin{split}
    i(t+1) = i(t) + \Delta \left( -m\ln{i(t)} + \frac{1+m\ln i(t)}{f^{2}} \right)i(t)- \gamma i(t)
\end{split}
\label{eqpowerlawlinearAI}
\end{equation}
Now in the very early stage of the epidemic, only a few agents in the population are infected, and hence $i(t) \sim 1/N$ where $N$ is the total number of agents in the population. This implies that during the early stage, the value of $\ln i(t)$ will not change much and will  be approximately a negative constant of the order of $\ln N$. We will denote $-\ln i(t)$ by $\alpha$ where $\alpha >0$.  Now, the critical value of the power $m$ to prevent the epidemic from spreading can be obtained from the condition $\frac{i(t+1)}{i(t)} < 1$, which gives,
\begin{equation}
    m^{crit}_{I} =\frac{\gamma f^{2} - \Delta}{\alpha \Delta\left(f^{2}-1\right)}\label{micrit}
\end{equation}
This implies that when the fraction of infected population taking adaptive action is related to the global prevalence data in a power-law manner, the disease will not spread in the population if $m < m^{crit}_{I}$. Fig.~\ref{fig:micrit} shows the variation of $m^{crit}_{I}$ with $f$ for a few different values of $\Delta$. The region above each curve represents the epidemic phase, while the region below the curve corresponds to the non-epidemic phase. Also, we can verify that in order to get a positive value for $m^{crit}_I$, the minimum possible value of $f$ is still given by Eq.~\ref{fmin_ai}.

Using Eq.~\ref{eqpowerlawlinearAI}, peak prevalence and epidemic size are given by Eq.~\ref{Imaxeqn} and Eq.~\ref{etaAI_r} respectively with, 
\begin{equation}
    \eta = m\alpha  + (1 - m \alpha)/f^{2} \label{em_fes}
\end{equation}
When the agents are static, we can obtain a lower bound for the critical value of $m$ say $m^{crit}_{I_-}$ by following a similar line of argument given in Sec.~\ref{constant_ai} leading to Eq.~\ref{perc2d}, which gives,
\begin{equation}
    m^{crit}_{I_-} = \dfrac{\log(\rho \pi b^2 -A_c) - \log(\rho \pi b^2) - \log(1-1/f^2)}{\alpha}\label{mperc}
\end{equation}
As in Sec.~\ref{constant_ai}, we expect this bound to hold for the AS and the ASI cases as well.

\subsubsection{Adaptive Susceptible (AS)}
Here, the fraction of susceptible agents $k_{S}$ following adaptation  is assumed to be $i(t)^{m}$. Although the evolution equation of the infected fraction is different from that of the AI case, it is easy to show that the critical value of $m$ is the same as given by Eq.~\ref{micrit}. On calculating the peak prevalence and final epidemic size, we get the same expressions as in AI model.

\subsubsection{Adaptive Infected and Susceptible (AIS)}
In this case, fractions of both infected and susceptible populations who adapt are related to the global prevalence in a power-law manner. For simplicity, we assume that the two fractions are the same $k_{I} = k_{S} = k_{IS} = i(t)^{m}$. 
Using the power-law form for $k_I$ and $k_S$ in Eq.~\ref{eqnASI} and neglecting terms of order $i(t)^2$ and higher, we get

\begin{equation}
\begin{split}
  i(t+1) &= i(t) \\
  & + s(t) \frac{\Delta}{f^{2}}  \left( i(t) + (f^{2} - 1)(i(t) -2i(t)^{m+1}+i(t)^{2m+1})\right)\\
  &-\gamma i(t) \label{mASIeqn}
\end{split}
\end{equation}
We consider values of $m$ between 0 and 1 and look for the behavior of Eq.~\ref{mASIeqn} in the lowest order in $m$ to find its critical value. However, unlike the AI and AS cases, the linear terms in $m$ now cancel out, and hence the lowest order term will have $m^2$. We use the expansions,
\begin{equation*}
    i(t)^{m+1} \approx i(t)\left(1 + m\ln(i(t))+\frac{m^{2}ln^{2}(i(t))}{2}\right)
\end{equation*}
\begin{equation*}
    i(t)^{2m+1} \approx i(t)\left(1 + 2m\ln(i(t))+\frac{4m^{2}ln^{2}(i(t))}{2}\right)
\end{equation*}
Using these and $\ln i(t) = -\alpha$, we can write Eq.~\ref{mASIeqn} as, 
\begin{equation}
    i(t+1) = i(t) + \frac{\Delta}{f^{2}}(1 + (f^{2} - 1)m^{2}\alpha^2)i(t) -\gamma i(t)
    \label{eqpowerlawlinarais}
\end{equation}
Thus the critical value of $m$ in this case is given by,
\begin{equation}
    m^{crit}_{IS} =\sqrt{
    \frac{\gamma f^{2} - \Delta}{\alpha^{2} \Delta \left(f^{2}-1\right)}} \label{mASIcrit}
\end{equation}
Comparing Eqs.~\ref{mASIcrit} with \ref{micrit}, we can see that the critical values of $m$ in the two cases are related by 
\begin{equation}
  m^{crit}_{IS} = \sqrt{m^{crit}_I/\alpha}  
\end{equation}
Using Eq.~\ref{eqpowerlawlinarais}, peak prevalence and final epidemic size follows from Eq.~\ref{Imaxeqn} and Eq.~\ref{etaAI_r} respectively with, 
\begin{equation}
    \eta^{m}_{AIS} = \frac{1 + (f^{2} - 1) m^{2}\alpha^{2}}{f^{2}}\label{metaAI}
\end{equation}
\subsubsection*{Effect of imperfect prevalence information}
A question of practical relevance is how epidemic dynamics are affected when agents respond to a perceived global prevalence level that deviates from the true prevalence $i(t)$. Such discrepancies can arise due to various factors, including inaccurate estimation of prevalence, limitations in data collection, or deliberate misinformation provided to the population. A simple way to capture the effect of such imperfect information is to assume that the true value of $i(t)$ is offset by a small amount $w$ where $|w|<<1$ during the the early stages of the epidemic.   In this case, we can write the fraction of agents who adapt as,  \\
\begin{equation}
    k = [i(t) \pm |w|]^{m}\label{info}
\end{equation}
where the $+|w|(-|w|)$ term corresponds to overestimation (underestimation). Now consider the AIS case in which both infected and susceptible agents adapt. An analysis similar to the one leading to Eq.~\ref{mASIcrit} shows that, in Eq.~\ref{mASIcrit}, $\alpha$ will now become $-\ln \left(i(t) \pm |w|\right)$, which implies that the critical value $m^{crit}_{SI}$ will now depend upon $w$. In particular, a positive value of $w$ will increase $m^{crit}_{SI}$ which implies that a slower initial response by the agents will now suffice to prevent the epidemic. The converse is true when $w$ is negative. 

\begin{figure}[!h]
    \centering
  \includegraphics[width=1\linewidth]{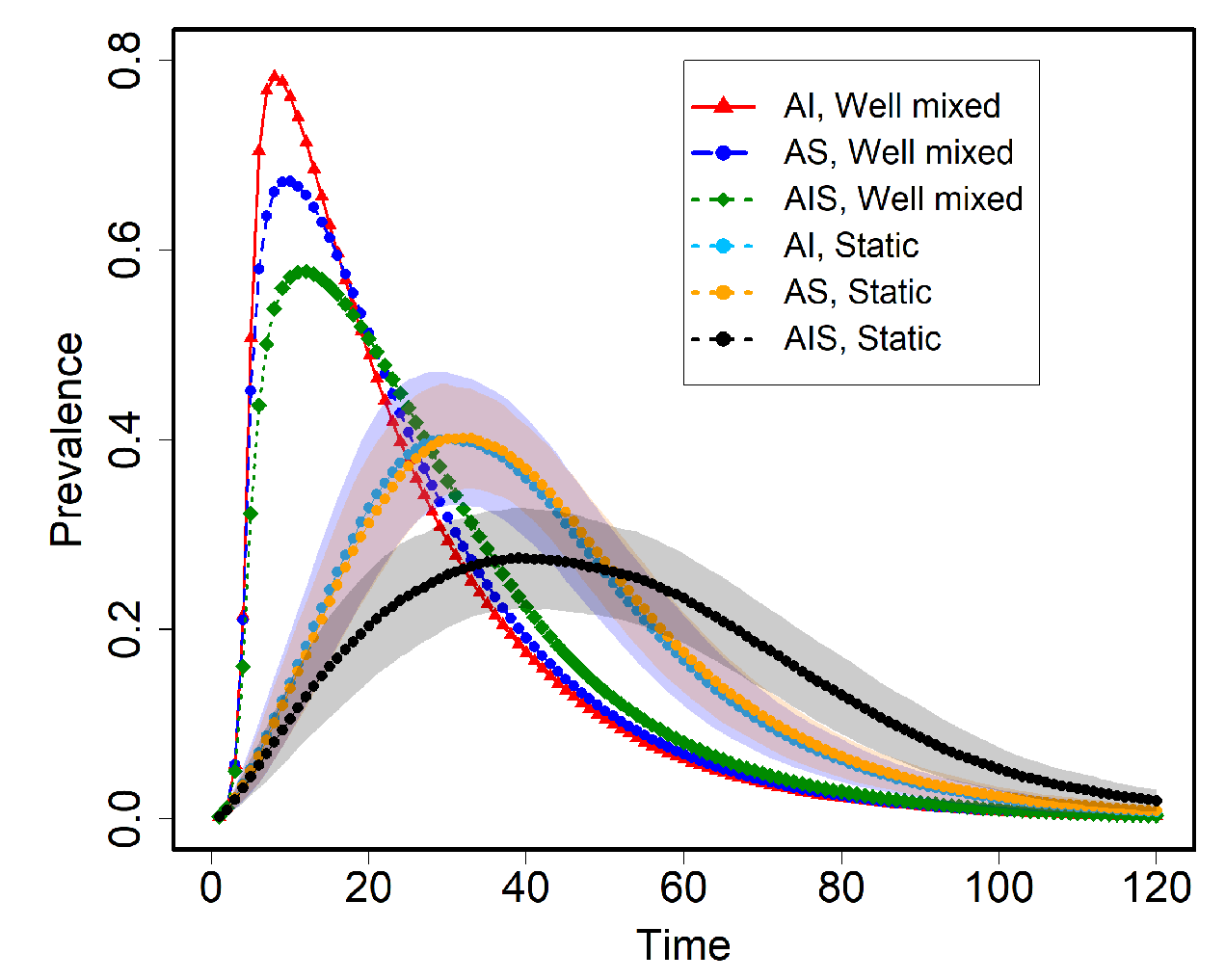}
    \caption{Typical variation of global prevalence with time in the AI, AS, and AIS cases when the fraction of agents who adapt is a power-law, $k = i^m$. For each case, prevalence
in a spatially well-mixed scenario (obtained by numerically
solving Eq.~\ref{eqnmi}, Eq.~\ref{mASIeqn}) is compared with that of spatially static
agents (obtained by monte carlo simulations). For the static case, the solid curves represent the average values obtained in $10^3$ trials and the shadows represent the spread.}
 \label{fig:AImpre2}
\end{figure}

\begin{figure}[!tbh]
    \centering
    \includegraphics[width=1\linewidth]{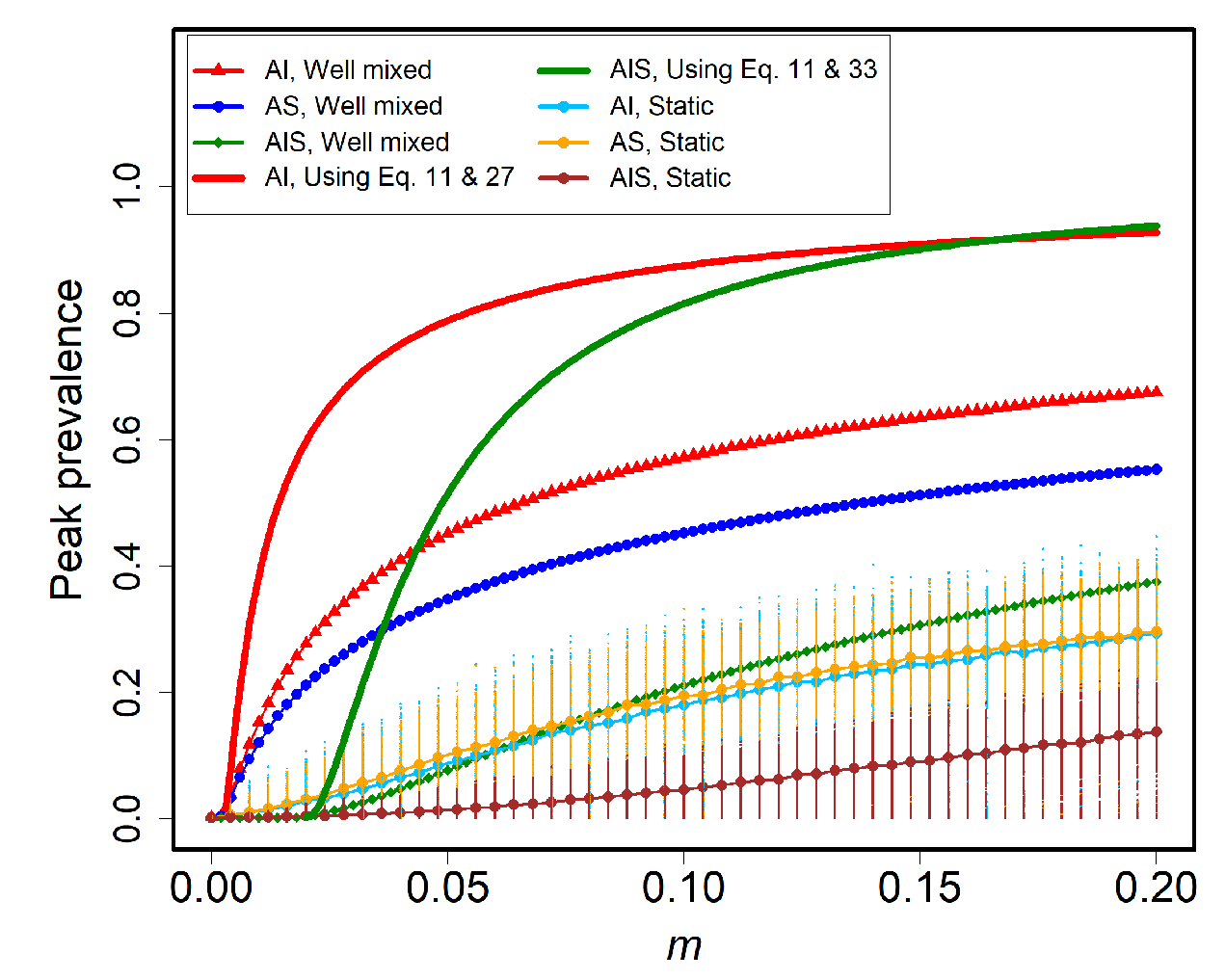}
    \caption{Variation of peak prevalence with $m$ for the AI, AS, and AIS scenarios. For the AI and AS cases, the transition point for the well-mixed case obtained from Eq.~\ref{micrit} is 
    $m^{crit}_{I} = m^{crit}_{S} \approx 0.0023$, and for AIS case, it is 
    $m^{crit}_{IS} \approx 0.0192$ (from Eq.~\ref{mASIcrit}). The lower bound for the three static cases obtained from Eq.~\ref{mperc} is $m^{crit}_{I-} \approx 0.076$.}
    \label{fig:mPPcombined2}
\end{figure}

\begin{figure}[!tbh]
    \centering
    \includegraphics[width=1\linewidth]{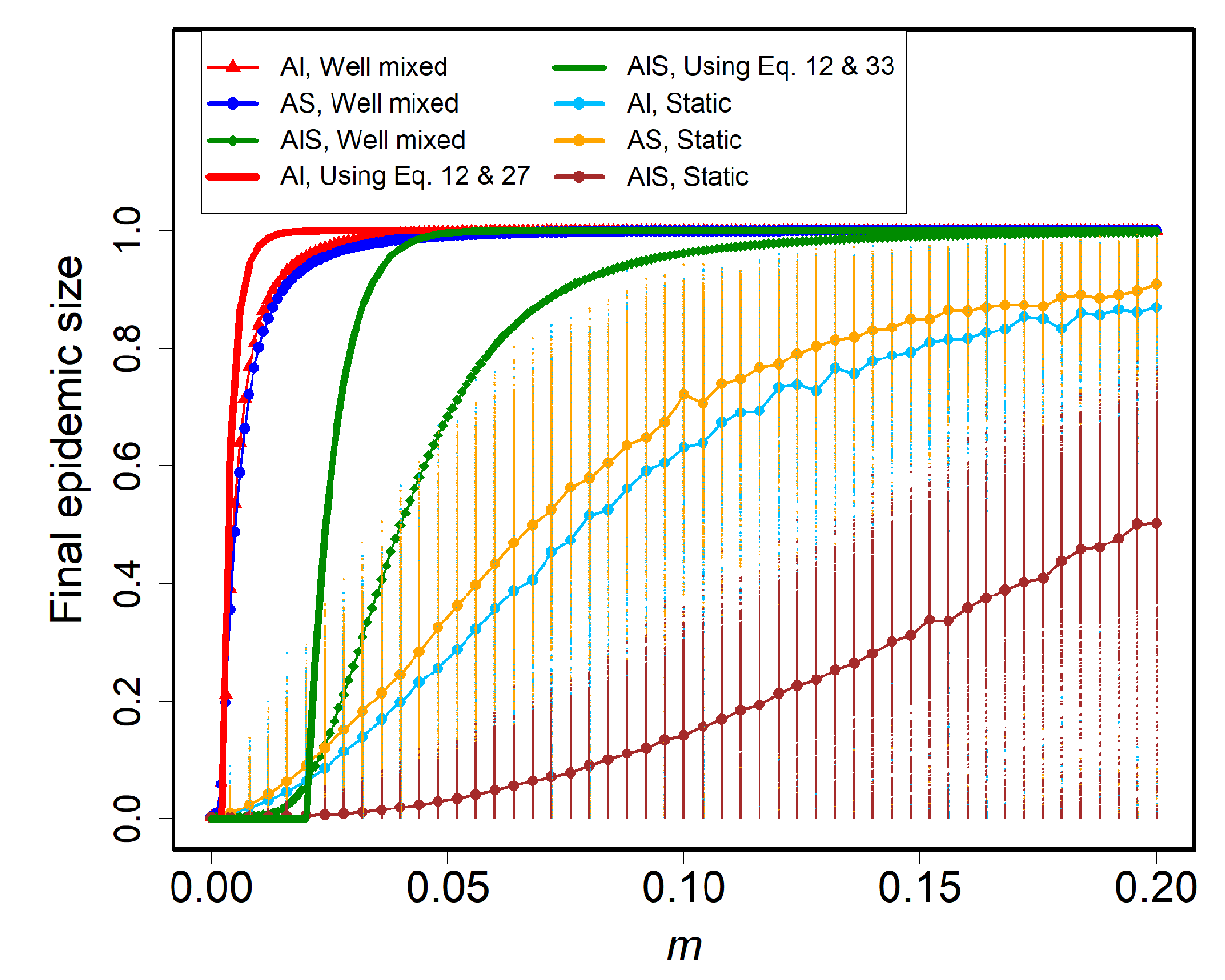}
    \caption{Variation of final epidemic size with $m$ for the AI, AS and AIS cases.}
\label{fig:mPPcombinedfe2}
\end{figure}

We now look into the dynamic behavior of the model with a power-law adaptation in detail and compare the results for AI, AS, and AIS cases. For the well-mixed scenario, the prevalence curves
are obtained by solving the evolution equations for $i(t)$ numerically and for the static case, the results
are obtained by monte carlo simulations.
Fig.~\ref{fig:AImpre2} shows typical prevalence curves for the AI, AS, and AIS cases. Fig.~\ref{fig:mPPcombined2} and Fig.~\ref{fig:mPPcombinedfe2} respectively show the variation of peak prevalence and final epidemic size with $m$. The peak prevalence and final epidemic size obtained using Eq.~\ref{Imaxeqn} and Eq.~\ref{etaAI_r} are shown for comparison. The threshold values obtained for the well-mixed case from Eq.~\ref{micrit} and Eq.~\ref{mASIcrit} are in good agreement with the numerical results. In the figures, we can see a significant reduction in the peak prevalence and final epidemic size with the AIS agents compared to the AI and AS cases for a given value of $m$.  Also, when $m$ is slightly above its critical value, the peak prevalence and the final epidemic size in the well-mixed cases show a much greater response to a change in $m$ compared to the static case. 


\section{Global prevalence based adaptation: Sigmoid-law}
\label{sec4}
\begin{figure}
    \centering
    \includegraphics[width=1\linewidth]{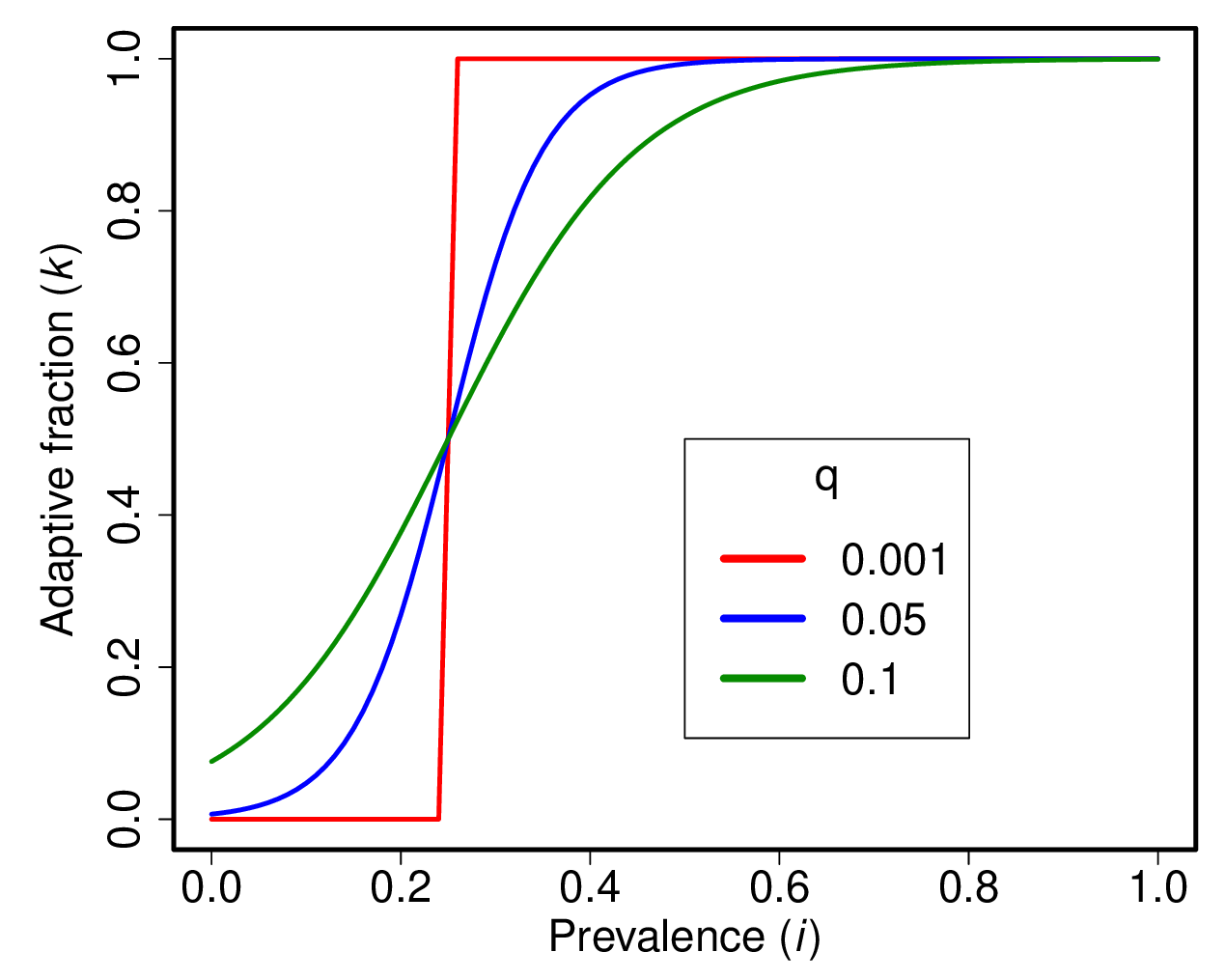}
    \caption{Variation of adaptive fraction $k$ with global prevalence $i$ for a fixed value of $p = 0.25$ and different values of $q$ in the sigmoid-law model of adaptation where $k =  \frac{1}{1 + e^{\frac{-(i - p )}{q}}}$.}
    \label{fig:siglaw}
\end{figure}

In the final adaptation scenario we consider, the fraction of infected or susceptible agents adapting is assumed to take the form of a sigmoid function of the global prevalence. i.e,
\begin{equation}
    k(t) = \frac{1}{1 + e^{\frac{-\left(i(t) - p \right)}{q}}}\label{Sigmoid}
\end{equation}
where $0<p<1$ and $0<q<<1$. Both $p$ and $q$ are parameters quantifying the nature of the response of the agents towards the global prevalence (See Fig.~\ref{fig:siglaw}). Here $p$ represents the level of global prevalence around which agents respond highly to the value of $i(t)$ and take adaptive measures. The parameter $q$ controls the width of the high response region around $p$. When $q << 1$, the sigmoid function approximates a threshold-based adaptation model, the value of the threshold being $p$. This can represent real-world interventions like prevalence-triggered lock-downs or other similar large-scale behavioral shifts.

\begin{figure}
    \centering
    \includegraphics[width=1\linewidth]{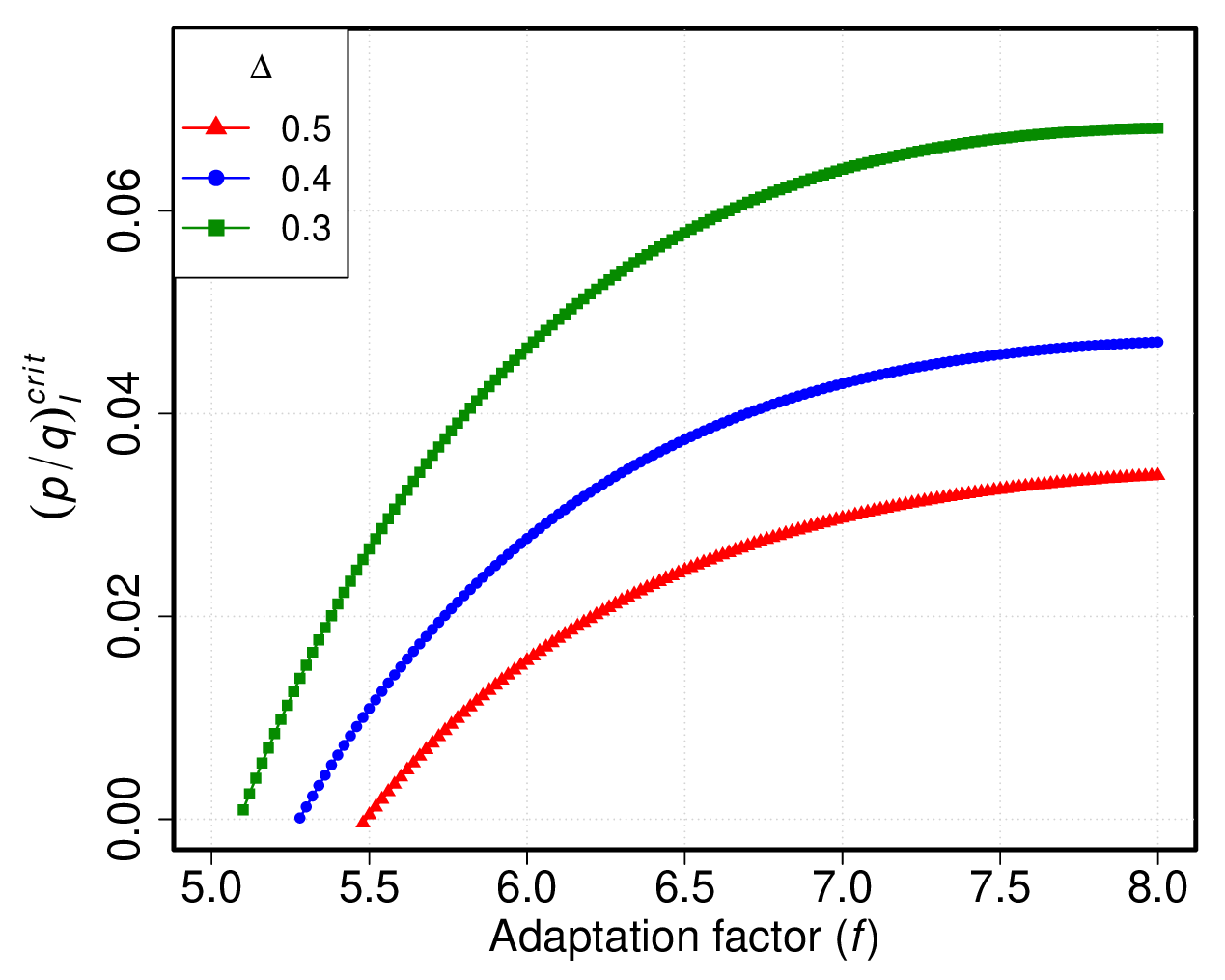}
    \caption{Variation of $(\frac{p}{q})_I^{crit}$ with adaptation factor $f$ for different values of $\Delta$. All $\Delta$ values chosen lie in the range $\gamma < \Delta \leq 2\gamma$.} 
    \label{pbyqcrit}
\end{figure}

It is easy to see that for a fixed value of $p$, the sigmoid function in Eq.~\ref{Sigmoid} when used in Eq.~\ref{eqn5} or Eq.~\ref{eqnASI} does not contribute any linear term in $i(t)$. In other words, in the initial phase of the epidemic, when $i(t) << 1$, we can set $i(t) = 0$ in Eq.~\ref{Sigmoid} and directly obtain the threshold condition. Hence, the critical values of $p/q$ for the AI, AS and AIS cases can be obtained by using Eq.~\ref{Sigmoid} with $i(t) = 0$ in Eq.~\ref{kIcrit} (for the AI and AS cases) and Eq.~\ref{KSIcrit} (for the AIS case). For AI and AS cases, we get the critical threshold, 
\begin{equation}
 \left(\frac{p}{q}\right)^{crit}_{I(S)} = \ln \left( \frac{\gamma f^{2} - \Delta}{f^{2} (\Delta - \gamma)} \right)\label{AIpbyqcrit} \end{equation}
For the AIS case, we get
\begin{equation}
 \left(\frac{p}{q}\right)^{crit}_{IS} = -\ln \left(\sqrt \frac{\Delta( f^{2} - 1)}{(\gamma f^{2} - \Delta)} - 1\right)\label{ASpbyqcrit}
\end{equation}
Thus only when $p/q$ is lower than these critical values, we expect a sigmoid-type response to prevent the spreading of the disease. The variation of $(p/q)_{I}^{crit}$ with $f$ for a few different values of $\Delta$ is shown in Fig.~\ref{pbyqcrit}.

Note that for the AI and AS cases, in Eq.~\ref{AIpbyqcrit}, the minimum value of $f$ to get a positive critical value is given by, 

\begin{equation}
    f_{min} = \sqrt{\dfrac{\Delta}{(2\gamma - \Delta)}}
\end{equation}
along with the condition $\Delta < 2\gamma$. Since $f_{min}$ has to be greater than one, we also require $\Delta > \gamma$. These conditions imply that sigmoid type adaptation is effective in the AI and AS cases only when $\gamma < \Delta \lessapprox 2 \gamma$. Now, from Eq.~\ref{ASpbyqcrit}, in the AIS case, the minimum value of $f$ to have a positive critical value is given by, 
\begin{equation}
  f_{min}  = \sqrt{\dfrac{3 \Delta}{(4\gamma - \Delta)}}
\end{equation}
along with the condition $\Delta < 4\gamma$. So the sigmoid type adaptation is effective in the AIS case only when $\gamma < \Delta \lessapprox 4 \gamma$.

For the AI and AS cases, peak prevalence and final epidemic size follows from Eq.~\ref{Imaxeqn} and Eq.~\ref{etaAI_r} respectively with,
\begin{equation}
   \eta =  \frac{1}{f^{2}}\left(1 + (f^{2}-1)\left(\frac{e^{\frac{p}{q}}}{1+e^{\frac{p}{q}}}\right)\right)
\end{equation}
and for the AIS case, with
\begin{equation}
   \eta =  \frac{1}{f^{2}}\left(1 + (f^{2}-1)\left(\frac{e^{\frac{p}{q}}}{1+e^{\frac{p}{q}}}\right)^{2}\right)
\end{equation}

When the agents are static, we can obtain a lower bound for the critical value of $p/q$ say $(p/q)^{crit}_{-}$ by substituting the sigmoidal form for the fraction in Eq.~\ref{perc2d}, which gives,
\begin{equation}
    (p/q)^{crit}_{-} = \log(A_c - \rho \pi b^2/f^2 ) - \log(\rho \pi b^2 - A_c) 
\end{equation}

\begin{figure}[!tbh]
    \centering
    \includegraphics[width=1\linewidth]{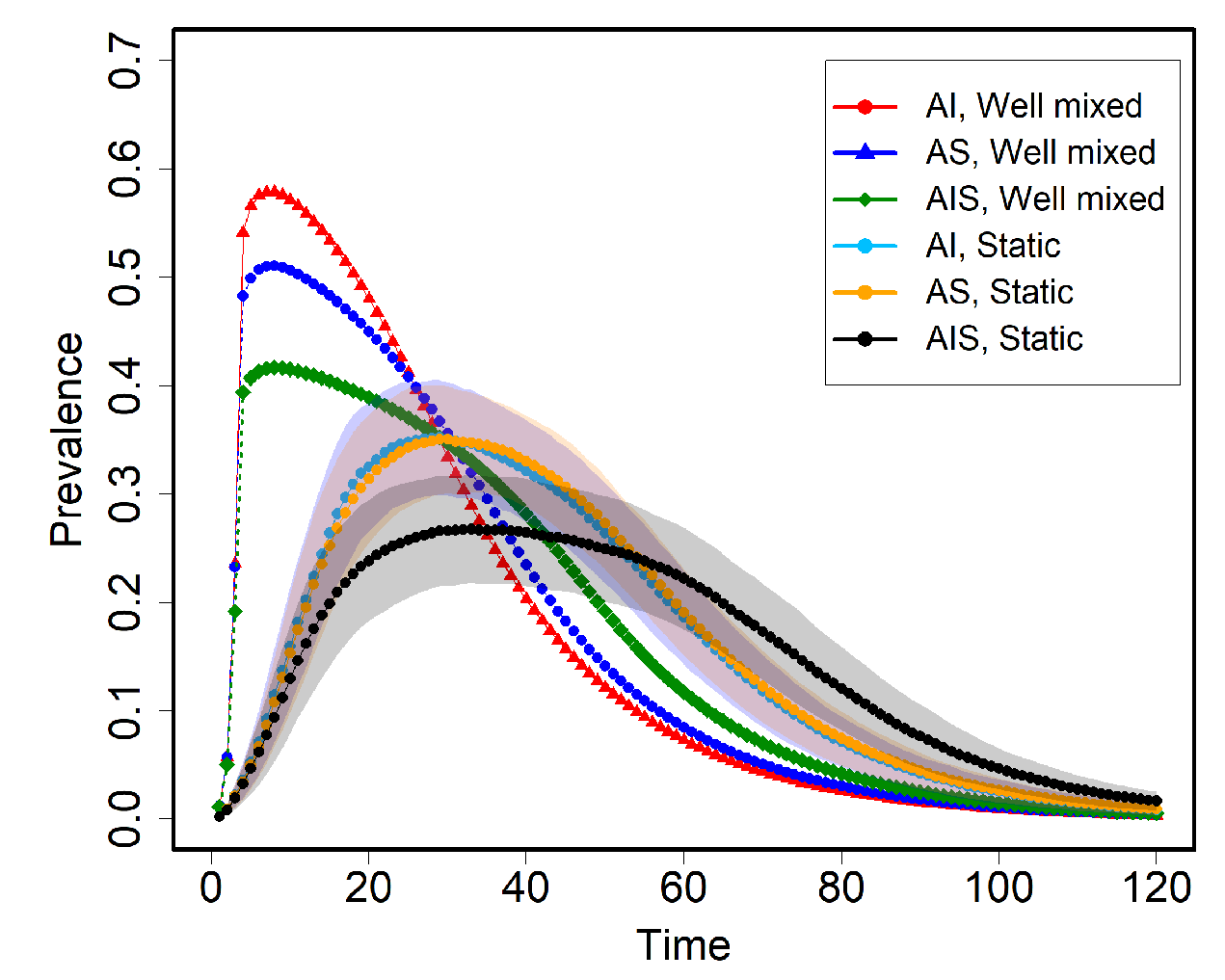}
    \caption{Variation of prevalence with time in the AI, AS, and AIS cases when the fraction of agents who
adapt follows a sigmoid-law $k =  \frac{1}{1 + e^{\frac{-(i - p )}{q}}}$. For each case, prevalence
in a spatially well-mixed scenario (obtained by numerically
solving Eq.~\ref{eqn2I}, Eq.~\ref{eqnS2} and Eq.~\ref{eqnASI} where $k$ is substituted with the sigmoid function) is compared with that of spatially static
agents (obtained by monte carlo simulations). The parameter values used are $p = 0.25$, $q = 0.1$.}
    \label{fig:q_Prevalence}
\end{figure}

\begin{figure}[!tbh]
    \centering
    \includegraphics[width=1\linewidth]{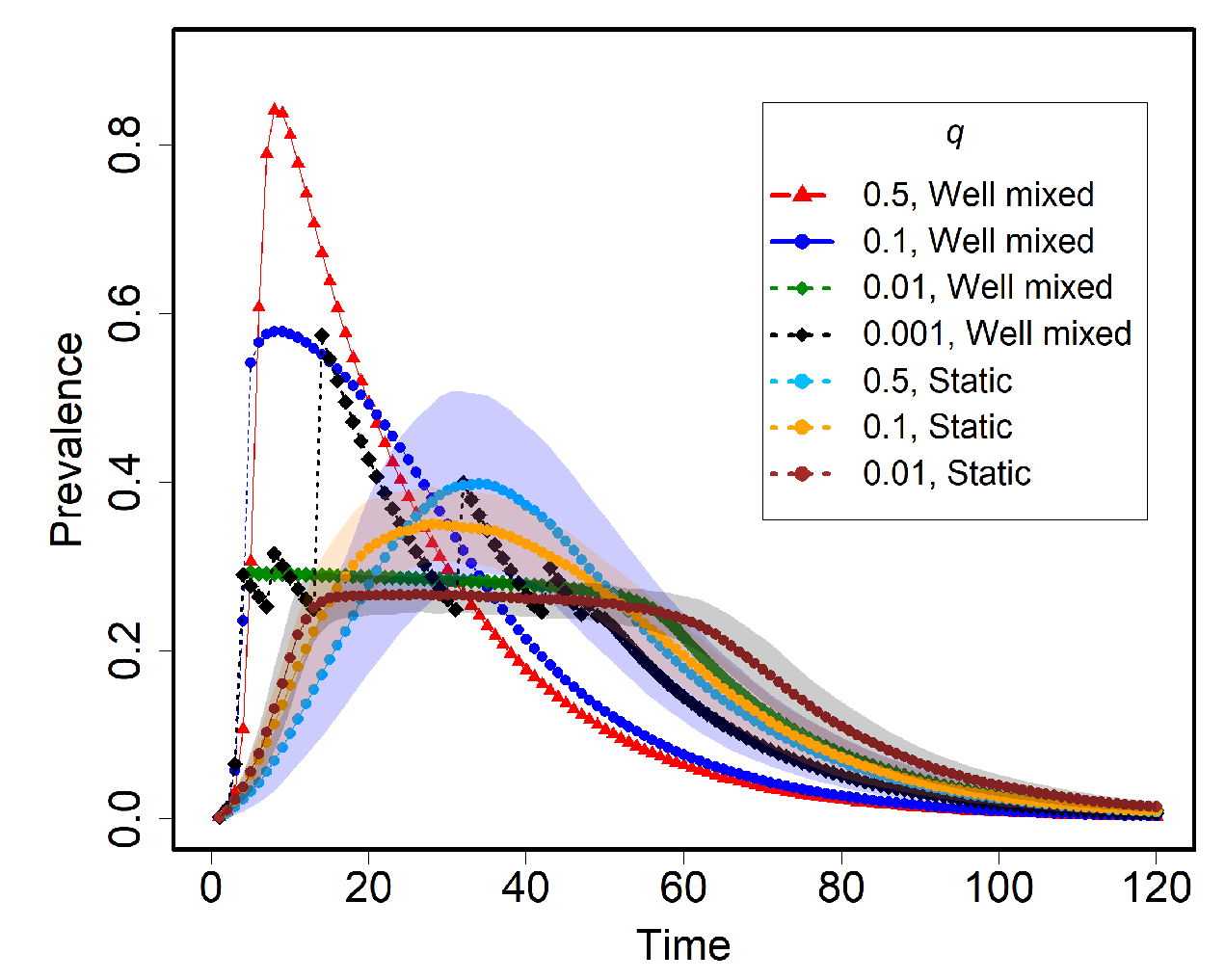}
    \caption{The variation of prevalence with time in the AI case when the fraction of agents who
adapt follows a sigmoid-law $k =  \frac{1}{1 + e^{\frac{-(i - p )}{q}}}$ showing the effect of varying the parameter $q$. For low enough $q$, adaptation leads to oscillations in prevalence.}
    \label{fig:AIsigmPre2}
\end{figure}

Fig.~\ref{fig:q_Prevalence} shows typical prevalence plots for the AI, AS, and AIS cases for the well-mixed and static scenarios. As the parameter $q$ affects the nature of the prevalence curves significantly, in Fig.~\ref{fig:AIsigmPre2}, we show prevalence plots for different values of $q$ in the AI case comparing the well-mixed and static scenarios. A notable feature here is that the prevalence shows oscillations for smaller $q$ values in the well-mixed case. The special limiting case of $q \rightarrow 0$ where the response curves in Fig.~\ref{fig:siglaw} become a step function was discussed in our earlier work~\cite{Akhil24}.   

\begin{figure}[!tbh]
    \centering
    \includegraphics[width=1\linewidth]{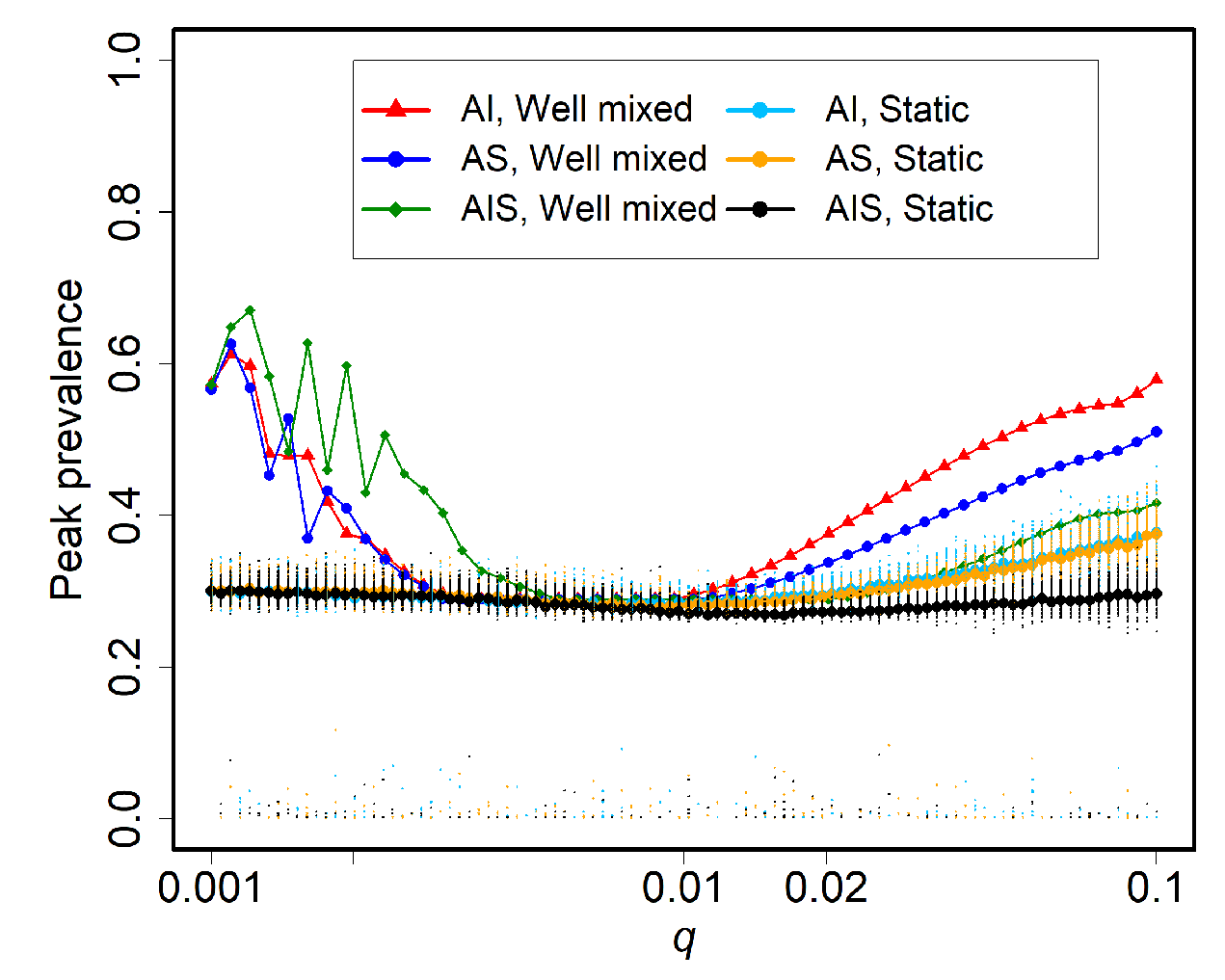}
    \caption{Variation of peak prevalence with $q$ for the  AI, AS and AIS cases with sigmoid-law adaptation. Here $p=0.25$.}
    \label{fig:q_pp25}
\end{figure}

\begin{figure}[!tbh]
    \centering
    \includegraphics[width=1\linewidth]{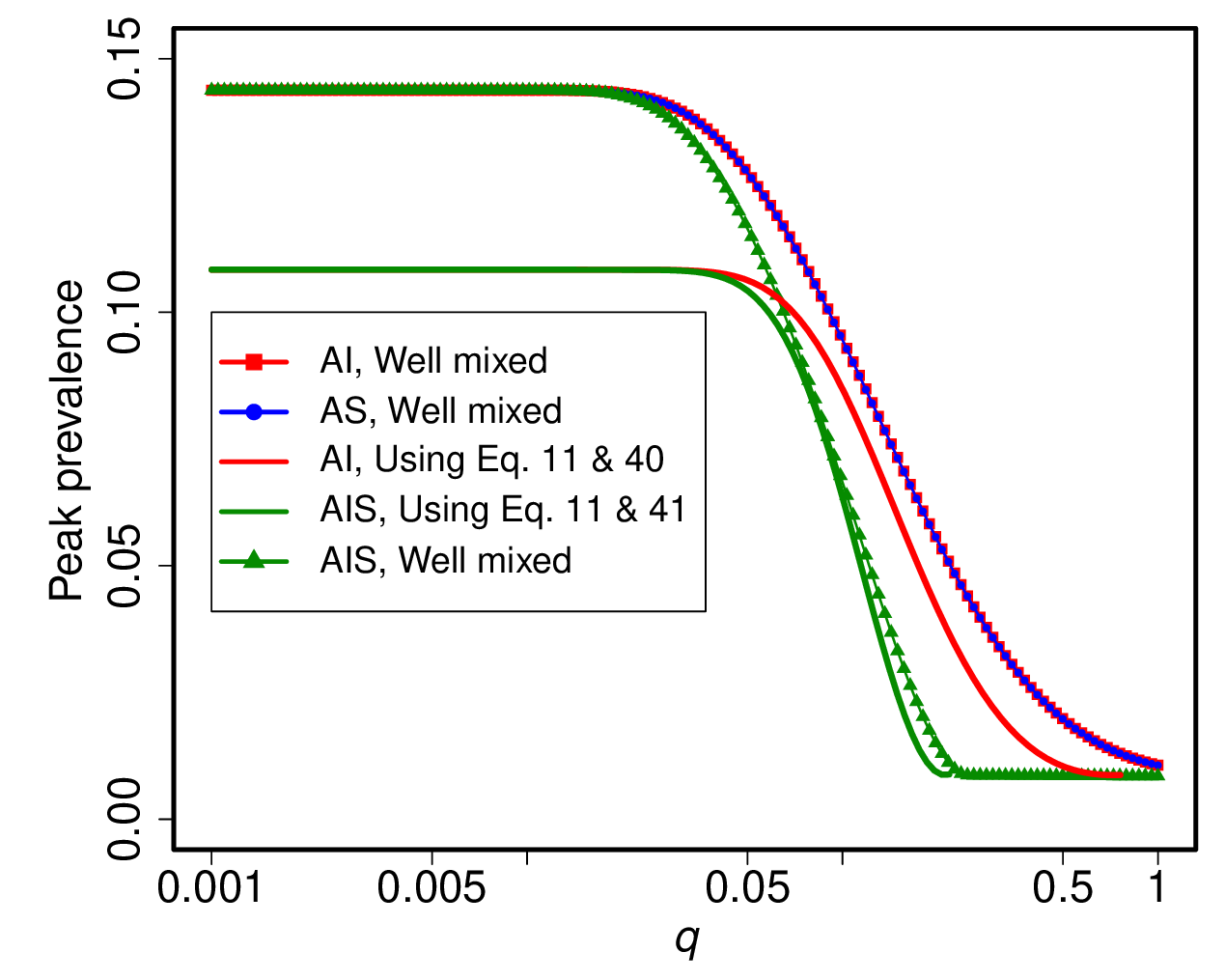}
    \caption{Variation of peak prevalence with $q$  for the AI, AS and AIS cases with sigmoid-law adaptation. Here we consider $\Delta = 0.08$. For the AI and AS cases, the transition
point for the well-mixed case obtained from Eq.~\ref{AIpbyqcrit} is $q^{crit}_{I} \approx q^{crit}_{S} = 0.76$, and for the AIS case, it is $q^{crit}_{IS} = 0.21$ (from Eq.~\ref{ASpbyqcrit}). The case with static agents do not show a transition for the $\Delta$ value considered and hence not shown.}
    \label{fig:q_pp2}
\end{figure}

\begin{figure}[!tbh]
    \centering
    \includegraphics[width=1\linewidth]{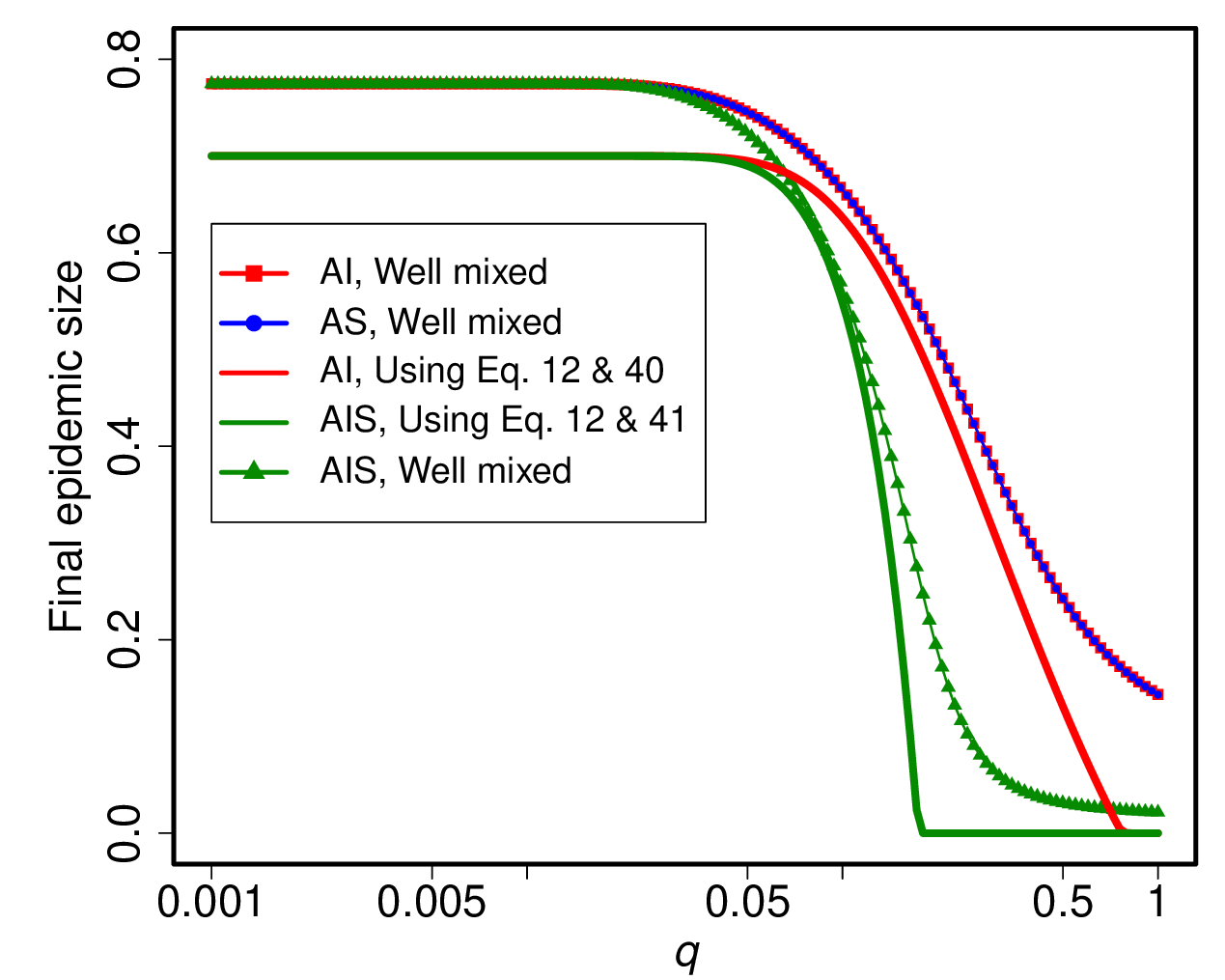}
    \caption{Variation of final epidemic size with $q$ for  AI, AS and AIS cases with $p = 0.25$. We use $\Delta = 0.08$. The case with static agents do not show a transition for the $\Delta$ value considered and hence not shown.}
    \label{fig:q_fes}
\end{figure}

Fig.~\ref{fig:q_pp25} shows the variation of peak prevalence with $q$ for the three cases AI, AS and AIS. Note that, here, as $(\Delta = 3.07) > (4\gamma = 0.2)$, sigmoid adaptation cannot prevent the epidemic from spreading. However, we can see that the well-mixed case shows a non-monotonic variation of peak prevalence with $q$ indicating that the disease is better suppressed for an intermediate range of values of $q$ for fixed values of other parameters. This is because when $q$ is extremely small, the agents adapt en masse when the prevalence crosses the threshold $p$ from below causing an initial suppression of the disease. However, when the threshold is crossed on the downside from above after a while, this results in a sudden removal of all adaptation and a further peaking of  prevalence (for example, see the curve corresponding to $q=0.001$ in Fig.~\ref{fig:AIsigmPre2}). At higher values of $q$, this effect disappears and the prevalence goes down after the initial peak itself. Oscillations in the peak prevalence can also be observed in Fig.~\ref{fig:q_pp25} for very small values of $q$ in the well-mixed case. However, we verified that this is mostly a numerical artifact seen  only when the initial infected fraction is $1/N$ and disappears for higher values. To clearly see the effect of sigmoid type adaptation, we have to consider values of $\Delta$ in the range $(\gamma, 4 \gamma)$. Fig.~\ref{fig:q_pp2} and Fig.~\ref{fig:q_fes} show the variation of peak prevalence and final epidemic size, respectively, with $q$  with $\Delta = 0.08$ and other parameter values same as in earlier cases. The transition point of  $q$ is compared with the critical values obtained from Eq.~\ref{AIpbyqcrit} and Eq.~\ref{ASpbyqcrit} with $p = 0.25$.  

\section{Conclusion}
\label{sec5}
In this work, we studied the impact of prevalence-based social adaptation by a population on the spatial spread of an epidemic. The social adaptation takes the form of reducing the characteristic range - the spatial extent over which disease transmission can occur - which is tantamount to adopting non-pharmaceutical measures like mask-wearing, social distancing or reducing interaction with others. We analysed three distinct adaptation schemes in which the fraction of agents who adapt is a constant, the adaptive fraction is dependent on global prevalence in a power-law manner, and the adaptive fraction is dependent on global prevalence in a sigmoid-law manner. Within each adaptation scenario, we further examined the cases of adaptation by infected agents, susceptible agents, and both types of agents. Our analysis considered two regimes representing the extremes of agent mobility: Spatially well-mixed regime where agents are repositioned randomly at each time step, enabling a mean-field treatment, and spatially static regime where agents remain fixed at the nodes of a random geometric graph, requiring simulation-based approaches. 

In the spatially well-mixed case, we obtained the critical fraction of agents in a population who must adapt in order to contain the epidemic in different adaptation scenarios. With spatially static agents, we established useful bounds for the critical adaptive fraction by drawing analogy with the continuum percolation of overlapping discs with a size distribution. Evolution of the epidemic is studied in detail by obtaining variation of prevalence with time, and variation of peak prevalence and size of the epidemic with adaptation parameters. In all case, comparison is done between well-mixed and static populations highlighting the influence of agent mobility on epidemic outcomes.

Our results show that an adapted fraction linear in the global prevalence perform similarly to a constant adaptive fraction in controlling the epidemic. In fact, for reasonable values of spatial density of agents and other parameters, we see that a superlinear dependence of adaptive fraction on global prevalence is necessary to effectively prevent the disease from spreading. With a sigmodial dependence of adaptive fraction on global prevalence, oscillations emerge in the prevalence when the width of the sigmoid is sufficiently narrow. Interestingly, the peak prevalence exhibits a non-monotonic dependence on the width of the sigmoid function, indicating that there is an optimal width that minimizes the epidemic's peak intensity.

The spatial adaptation framework developed here is general and can be extended to explore other distance-dependent spreading processes where a fraction of the population engages in adaptive behavior. While this study focused on the SIR model, our approach can be readily applied to other similar epidemiological models like Susceptible-Infected (SI) for persistent infections and Susceptible-Exposed-Infected-Recovered (SEIR) for diseases with incubation periods. Introducing distributions for the adaptation factor and other parameters of the model could capture more realistic variations in individual behavior. Future works could explore the role of agent mobility beyond the well-mixed and static limits considered here, incorporating correlated motion, home-based movements, or localized diffusion and it will useful to see how different models for agent mobility affect the findings.

 \subsection*{Acknowledgments}
Authors acknowledge the use of the high-performance computing cluster established at Cochin University of Science and Technology (CUSAT) under the Rashtriya Uchchatar Shiksha Abhiyan (RUSA 2.0) scheme (No. CUSAT/PL(UGC).A1/2314/2023, No: T3A).

\end{document}